\documentclass[aps,pra,superscriptaddress,amsmath,amssymb,showpacs,twocolumn,notitlepage]{revtex4-1}
\usepackage{amsmath,amssymb,amsfonts,latexsym,color,epsfig,textcomp}
\usepackage{subfigure}
\usepackage[latin1]{inputenc}
\usepackage{dcolumn}
\usepackage{bm}
\usepackage{graphicx}
\usepackage[english]{babel}
\usepackage{booktabs}
\usepackage{multirow}
\usepackage[table,xcdraw]{xcolor}

\DeclareMathOperator{\Rea}{Re}

\newcommand{\g}[1]{\displaystyle{\gamma_#1}}
\newcommand{\gp}[1]{\displaystyle{\gamma'_#1}}
\newcommand{\al}{\displaystyle{\alpha}}
\def\bbm[#1]{\mbox{\boldmath $#1$}}

\newcommand{\rr}{{\bf r}}
\newcommand{\de}{\delta}
\newcommand{\et}{\eta}

\begin{document}

\title{Light-induced optomechanical forces in graphene waveguides}

\author{Brahim Guizal}
\affiliation{Laboratoire Charles Coulomb (L2C), UMR 5221 CNRS-Universit\'{e} de Montpellier, F- 34095 Montpellier, France}
\author{Mauro Antezza}
\email{Correspondance to: mauro.antezza@umontpellier.fr}
\affiliation{Laboratoire Charles Coulomb (L2C), UMR 5221 CNRS-Universit\'{e} de Montpellier, F- 34095 Montpellier, France}
\affiliation{Institut Universitaire de France - 1 rue Descartes, F-75231 Paris, France}

\date{\today}

\begin{abstract}
We show that the electromagnetic forces generated by the excitations of a mode in graphene-based optomechanical systems are highly tunable by varying the graphene chemical potential, and orders of magnitude stronger than usual non-graphene-based devices, in both attractive and repulsive regimes. We analyze coupled waveguides made of two parallel graphene sheets, either suspended or supported by dielectric slabs, and study the interplay between the light-induced force and the Casimir-Lifshitz interaction. These findings pave the way to advanced possibilities of control and fast modulation for optomechanical devices and sensors at the nano- and micro-scales.
\end{abstract}



\maketitle

\section{Introduction}\label{sec:intro}

The electromagnetic field may induce forces on bodies trough several mechanisms. One of them is the omnipresent fluctuation-induced attractive Casimir-Lifshitz (CL) or van der Waals interactions dominating at sub-micron bodies's separations  \cite{CasimirBook}, with destructive effects in nano- and micro- electromechanical devices  \cite{Chan2001}. If an electromagnetic mode is excited in the system by an external source, it produces an extra light-induced (LI) force  \cite{Povinelli2005,Riboli2008}, which can be attractive or repulsive, possibly overcoming/balancing the CL force. To design LI forces, several materials and complex nano-structured geometries have been intensively studied (photonic crystals, resonators, metamaterials), mainly to maximize their repulsion or to increase the optical interactions, hence improving actuations and functionalities in nano-opto(electro)-mechanical systems (NOEMS) and sensors  \cite{ReviewCapasso, PRL_ANTEZZA}. In particular, the mechanism for increasing the LI interaction calls for a subtle interplay between strong confinements of the fields and their spatial oscillations  \cite{Oskooi2011}, together with  a reduction of the group velocity  \cite{Povinelli2005}. Metals strongly confine the fields, but the LI force is limited by losses  \cite{Wolf2009} and the repulsion is contrasted by huge CL forces. Dielectrics have a weaker CL force, they confine less efficiently the fields, but once nano-structured they have ultra-low group velocities  \cite{Rodriguez2011}. They represent an optimal compromise, allowing the largest values of repulsion  \cite{Oskooi2011}, orders of magnitude higher than non-structured dielectrics.  

Here we propose the exploitation of graphene sheets  \cite{CastroRMP2008} in optomechanical waveguides systems to manage LI interactions. Remarkably, graphene manifests low group velocity modes, a strong metallic ability to confine them, it is practically lossless in a wide region of frequencies, and gives rise to very weak CL forces.  Furthermore, the LI force becomes tunable by varying the graphene Fermi level via an electrostatic voltage or via chemical doping.  These unique features make graphene sheets able to strongly increase the repulsion, up to 1-2 orders of magnitude higher than the best nano-structured systems. Such electromagnetic properties are combined with peculiar mechanical properties (low density and bending stiffness, large modulus of elasticity) making them attractive for optomechanics  \cite{Wang2014}.

In section \ref{sec:system} we describe the physical systems, in section \ref{sec:DispRel} we derive the dispersion relations, in section \ref{sec:SlaGra} we discuss the optical properties of  silicon and graphene, in section \ref{sec:Length} we analyze the typical length scales required to use the lossless assumption, in section \ref{sec:press} we derive the LI pressure, in section \ref{sec:Casimir} we derive the CL pressure, in section \ref{sec:numres} we discuss the numerical results for the LI and CP pressures, and finally in section \ref{sec:conclusions} we provide the conclusions and perspectives.

\begin{figure}[htb]
\includegraphics[width=0.47\textwidth]{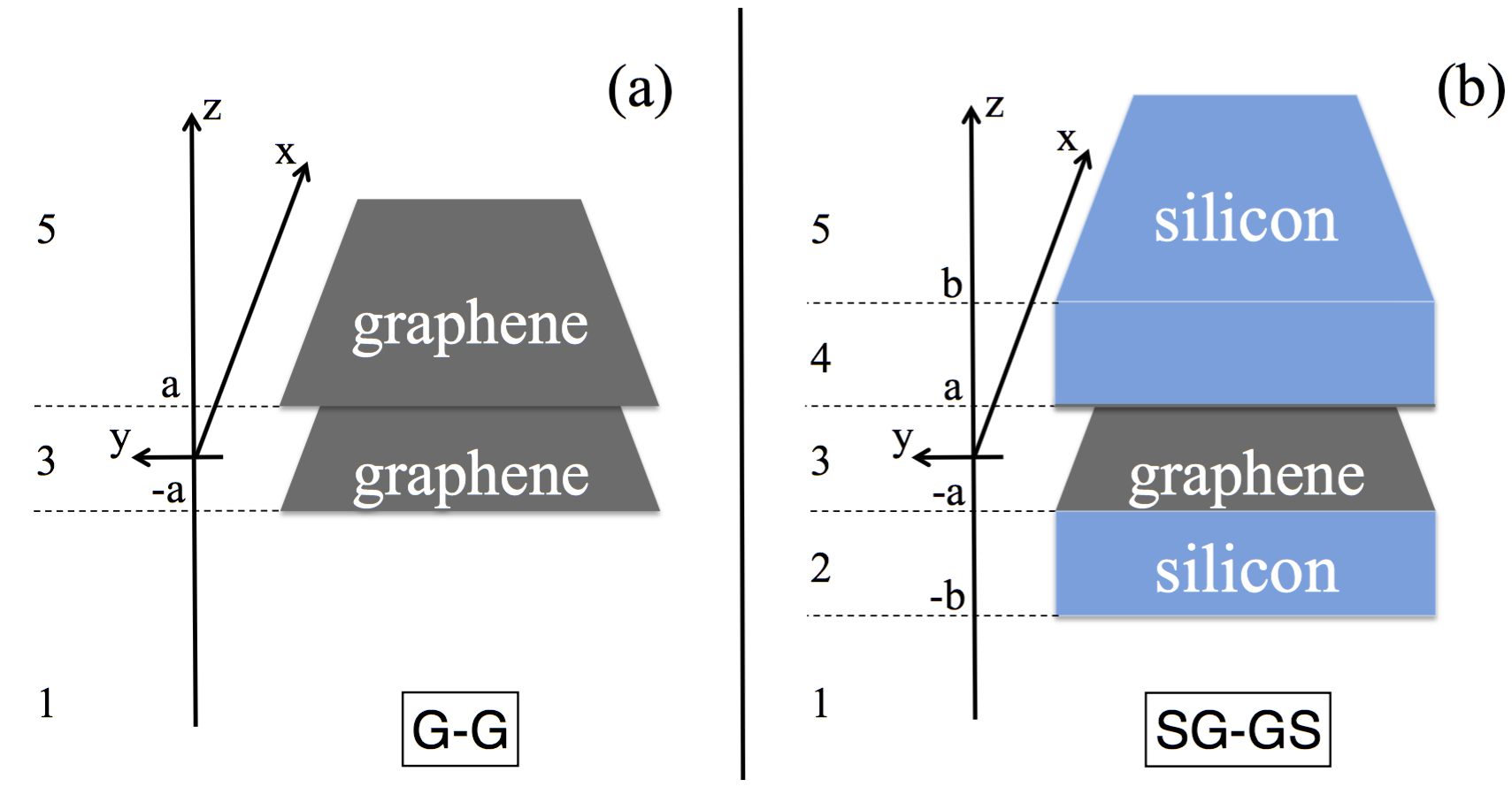}
\caption{\label{fig:schema}\footnotesize (color online). Scheme of the coupled waveguides system. Panel (a): the suspended graphene-graphene configuration (G-G). Panel (b): the slab-supported graphene-graphene configuration (SG-GS). The distance between the graphene sheets is $2a$, the thickness of the supporting Silicon slabs is $s=b-a$.}
\end{figure}

\section{Physical system}\label{sec:system}
We consider the interaction between two types of planar parallel coupled waveguides: a first  configuration is made by two suspended graphene sheets (G-G) at a distance $2a$ from each other [see Fig.\ref{fig:schema}(a)], and orthogonal to the $z$ axis. The second (SG-GS) consists of two graphene sheets, each one supported by a slab of thickness $s$ [see Fig.\ref{fig:schema}(b)]. Graphene and slabs are characterized by the conductivity $\sigma(\omega)=\sigma_\textrm{R}(\omega)+i\sigma_\textrm{I}(\omega)$ and the relative dielectric permittivity $\varepsilon(\omega)=\varepsilon_\textrm{R}(\omega)+i \varepsilon_\textrm{I}(\omega)$ [regions 2 and 4 in Figure \ref{fig:schema}(b)], respectively. Extension to configurations with non-identical graphene sheets or non identical slabs can be done straightforwardly. The external and central regions (regions 1, 3 and 5) are not filled by any materials ($\varepsilon=1$). LI modes are assumed to be excited and propagate in the $x$ direction, at frequency $\omega$, $y$ being the direction of invariance.

Electromagnetic forces (both CL and LI) acting on any of the two waveguides can be calculated by  \cite{Jackson,LL} ${\bf F}=\int_{\Sigma} {\bf T}({\bf r}) \cdot {\bf n}\;d\sigma$, where  $\Sigma$ is a closed oriented surface in vacuum enclosing the object and ${\bf T}=\langle\mathbb{T}({\bf r},t)\rangle_t$ is the time averaged Maxwell stress tensor. The CL force is not monochromatic, losses cannot be neglected, and it can be expressed as a sum over all available modes in the systems populated  by both vacuum and thermal field fluctuations. If the waveguides are close enough (but not too close to form a graphene bilayer) one can safely approximate the CL pressure with that occurring between infinite planes  \cite{CasimirBook,Antezza2011,variCasimirGraphene}.  The LI force is monochromatic, hence it is possible to further simplify the problem by choosing frequencies where the system is lossless, allowing for direct analytical expressions of the pressure, which now reduces to its $z$ component  \cite{Riboli2008}  (see Appendix \ref{sec:PhySysEleFor} for more details)
\begin{multline}\label{eq:LIpress}
p_{\textrm{LI}}=\frac{\varepsilon_0}{4}\left[|E_x|^2+|E_y|^2-|E_z|^2+\right.\\
\left.\mu_0^2c^2\left(|H_x|^2+|H_y|^2-|H_z|^2\right)\right],
\end{multline} 
 to be evaluated in the region between the two waveguides. Here we assumed that negative (positive) force corresponds to attraction (repulsion).

\section{Dispersion Relations}\label{sec:DispRel}
In order to derive the dispersion relations, i.e. $\al(\omega)$, for the  TE/TM symmetric/antisymmetric (s/a) modes we use the solution of the Maxwell equations in the different homogenous media of the structure, and impose the boundary conditions.

The invariance of the structures in the $y$ direction allows to classify the field modes in two different polarization states: the Transverse Electric (TE) and the Transverse Magnetic (TM). The TE polarization is characterized by $E_x=E_z=H_y=0$, and since 
\begin{equation}\label{eq:maxH}
H_z=-\frac{i}{\omega \mu_0}\partial_xE_y \;\;\; \text{and} \;\;\;H_x=\frac{i}{\omega \mu_0}\partial_zE_y, 
\end{equation}
the electromagnetic field can be completely determined by the knowledge of $E_y(x,z)$.  The TM polarization is characterized by $H_x=H_z=E_y=0$, and since 
\begin{equation}\label{eq:maxE}
E_z=\frac{i}{\omega \varepsilon_0 \varepsilon}\partial_xH_y\;\;\; \text{and} \;\;\;E_x=-\frac{i}{\omega \varepsilon_0 \varepsilon}\partial_zH_y,
\end{equation}
the electromagnetic field can be completely determined by the knowledge of $H_y(x,z)$. Each of the TE and TM modes can be further classified as symmetric or antisymmetric depending on the symmetry properties of the field (more precisely of $E_y(x,z)$ for TE modes and of $H_y(x,z)$ for TM modes) with respect to the $z=0$ plane.

We introduce a general field $\Phi$ in the five regions of space, which will be $\Phi\equiv E_y$ for the TE modes, and $\Phi\equiv H_y$ for the TM modes: 

\begin{widetext}
\begin{eqnarray}\label{eq:fields}
\begin{array}{ll}
\Phi_1(x,z)=A_1\;e^{\g{1}(z+b)}\;e^{i\al x}, \;\;\; &  \textrm{if} \;\;\; z\leq -b;\\
\\
\Phi_2(x,z)=\left[A_2\;e^{i\g{2}(z+a)}+B_2\;e^{-i\g{2}(z+a)}\right]\;e^{i\al x}, \;\;\;& \textrm{if} \;\;\; -b\leq z\leq -a;\\
\\
\Phi_3(x,z)=\left[A_3\;\cosh{(\g{1}z)}+B_3\;\sinh{(\g{1}z)}\right]\;e^{i\al x}, \;\;\;& \textrm{if} \;\;\; -a \leq z\leq a;\\
\\
\Phi_4(x,z)=\left[A_4\;e^{i\g{2}(z-a)}+B_4\;e^{-i\g{2}(z-a)}\right]\;e^{i\al x}, \;\;\;& \textrm{if} \;\;\; a \leq z \leq b;\\
\\
\Phi_5(x,z)=A_5\;e^{-\g{1}(z-b)}\;e^{i\al x}, \;\;\; &  \textrm{if} \;\;\; b\leq z;\\
\end{array}
\end{eqnarray}
\end{widetext}
where $\al(\omega)$ is the propagation constant along $x$, $\g{1}(\omega)=\sqrt{\al^2-k^2}$, $\g{2}(\omega)=\sqrt{k^2\varepsilon-\al^2}$, and $k=\omega/c$.
In general $\g{i}$ (with $i=1,2$) and $\al$ are complex quantities. We impose now the boundary conditions at the four interfaces, and solve the resulting linear system for the field coefficients $A_i$ and $B_i$ appearing in (\ref{eq:fields}).

\subsection{TE Modes}\label{ssec:TEs}
For the TE modes,  $E_y\equiv\Phi$ is continuous at the four interfaces, while $H_x$ [given by (\ref{eq:maxH})] is continuous at interfaces without graphene, and experiences a jump equal to the surface  current density $J_{x}=\sigma E_{y}(x,z)$ at the interfaces with graphene: 

\begin{eqnarray}\label{eq:clTE}
z=-b:&&
\left\{ 
\begin{array}{ll}
E_{2y}(x,-b)=E_{1y}(x,-b)\\
H_{2x}(x,-b)=H_{1x}(x,-b)
\end{array}
\right.\\
z=-a:&&
\left\{ 
\begin{array}{ll}
E_{3y}(x,-a)=E_{2y}(x,-a)\\
H_{3x}(x,-a)-H_{2x}(x,-a)=\sigma E_{2y}(x,-a)
\end{array}
\right.\\
z=a:&&
\left\{ 
\begin{array}{ll}
E_{4y}(x,a)=E_{3y}(x,a)\\
H_{4x}(x,a)-H_{3x}(x,a)=\sigma E_{4y}(x,a)
\end{array}
\right.\\
z=b:&&
\left\{ 
\begin{array}{ll}
E_{5y}(x,b)=E_{4y}(x,b)\\
H_{5x}(x,b)=H_{4x}(x,b)
\end{array}
\right.
\end{eqnarray}
For the symmetric (antisymmetric) mode, we set $B_3=0$ ($A_3=0$) and find $A_1=A_5$,  $A_2=B_4$, and $A_4=B_2$ ($A_1=-A_5$,  $A_2=-B_4$, and $A_4=-B_2$). Then, by elimination of the coefficients we obtain the dispersion relation for the TE symmetric and antisymmetric modes:
\begin{multline}\label{eq:dispTE}
\phi\;(\g{1}+i\g{2})\left[i(\g{2}-\et)+\g{1}\;F(\g{1}a)\right]\\
+\phi^{-1}\;(\g{1}-i\g{2})\left[i(\g{2}+\et)-\g{1}\;F(\g{1}a)\right]=0,
\end{multline}
where $\phi=e^{is\g{2}}$, $s=b-a$, $\et=\sigma k Z_0$  ($Z_0=\sqrt{\mu_0/\varepsilon_0}$ being the impedance of vacuum), and where we introduced the function:
\begin{equation}
F(x)=\left\{
\begin{array}{ll}
\tanh(x) & \textrm{for the symmetric mode},\\
\coth(x) & \textrm{for the antisymmetric mode}.
\end{array}
\right.
\end{equation}
We note that Eq.~\eqref{eq:dispTE} has a first solution $\g{2}=0$, that once substituted in (\ref{eq:fields}) implies a zero electromagnetic field everywhere. Hence we can exclude this solution and assume that $\g{2}\neq0$.

 \subsection{TM Modes}\label{ssec:TMs}
For the TM modes,  $H_y\equiv\Phi$ is continuous at the  interfaces without graphene, and experiences a jump equal to the opposite of the surface current density $-J_{y}=-\sigma E_{x}(x,z)$ at the interfaces with graphene, while $E_x$ [given by (\ref{eq:maxE})] is continuous at the four interfaces: 

\begin{eqnarray}\label{eq:clTM}
z=-b:&&
\left\{ 
\begin{array}{ll}
E_{2x}(x,-b)=E_{1x}(x,-b)\\
H_{2y}(x,-b)=H_{1y}(x,-b)
\end{array}
\right.\\
z=-a:&&
\left\{ 
\begin{array}{ll}
E_{3x}(x,-a)=E_{2x}(x,-a)\\
H_{2y}(x,-a)-H_{3y}(x,-a)=\sigma E_{2x}(x,-a)
\end{array}
\right.\\
z=a:&&
\left\{ 
\begin{array}{ll}
E_{4x}(x,a)=E_{3x}(x,a)\\
H_{3y}(x,a)-H_{4y}(x,a)=\sigma E_{4x}(x,a)
\end{array}
\right.\\
z=b:&&
\left\{ 
\begin{array}{ll}
E_{5x}(x,b)=E_{4x}(x,b)\\
H_{5y}(x,b)=H_{4y}(x,b)
\end{array}
\right.
\end{eqnarray}
For the TM symmetric (antisymmetric) mode, we set $B_3=0$ ($A_3=0$) and find, as for the TE mode, $A_1=A_5$,  $A_2=B_4$, and $A_4=B_2$ ($A_1=-A_5$,  $A_2=-B_4$, and $A_4=-B_2$). Then, by elimination of the coefficients we obtain the dispersion relation for the TM symmetric and antisymmetric modes:
\begin{multline}\label{eq:dispTM}
\phi\;(\g{1}+i\gp{2})\left[i\gp{2}+\g{1}\;(1-\de)\;F(\g{1}a)\right]+\\
\phi^{-1}\;(\g{1}-i\gp{2})\left[i\gp{2}-\g{1}\;(1+\de)\;F(\g{1}a)\right]=0,
\end{multline}
where $\phi$, $s$, $F(x)$ are the same as for the TE dispersion equation (\ref{eq:dispTE}), while $\gp{2}=\g{2}/\varepsilon$ and $\de=\sigma Z_0 \gp{2}/k$. 

We note that Eq.~\eqref{eq:dispTM} has a first solution $\g{2}=0=\gp{2}$, that once substituted in (\ref{eq:fields}) implies a zero electromagnetic field everywhere. Hence we can exclude this solution and assume that $\g{2}\neq0\neq\gp{2}$.

\begin{center}
\begin{table*}[htb]
\centering
\label{tab_eq}
\begin{tabular}{|l|l|l|l|l|l|l|}
\hline
\multirow{2}{*}{}              & \multicolumn{3}{c|}{TE} & \multicolumn{3}{c|}{TM} \\ \cline{2-7} 
                               &  \multicolumn{1}{c|}{S-S}      &     \multicolumn{1}{c}{G-G}     &  \multicolumn{1}{|c|}{SG-GS}       &     \multicolumn{1}{c|}{S-S}      &     \multicolumn{1}{c}{G-G}     &  \multicolumn{1}{|c|}{SG-GS}      \\ \hline
\multicolumn{1}{|c|}{Region 2} &   \multicolumn{1}{c|}{ Eq.  (\ref{eq:TEreg2}) or (\ref{eq:TEreg2form2})}     &  \multicolumn{1}{c}{no modes}      &  \multicolumn{1}{|c|}{ Eq.  (\ref{eq:TEreg2}) or (\ref{eq:TEreg2form2})}       &    \multicolumn{1}{c|}{ Eq.  (\ref{eq:TMreg2}) or (\ref{eq:TMreg2form2})}      &     \multicolumn{1}{c}{no modes}   &    \multicolumn{1}{|c|}{ Eq. (\ref{eq:TMreg2}) or (\ref{eq:TMreg2form2})}    \\ \hline
\multicolumn{1}{|c|}{Region 3}        &  \multicolumn{1}{c|}{no modes}     &   \multicolumn{1}{c}{no modes}     &   \multicolumn{1}{|c|}{no modes}    &   \multicolumn{1}{c|}{no modes}     &      \multicolumn{1}{c}{ Eq.  (\ref{eq:ggr3TM})}    &    \multicolumn{1}{|c|}{ Eq.  (\ref{eq:TMu3})}    \\ \hline
\end{tabular}
\caption{\label{Tavola}Equation for the TE and TM modes dispersion relations in the lossless case, corresponding to the slab-slab (S-S), graphene-graphene (G-G), and slabs supported graphene-graphene (SG-GS) configurations.}
\end{table*}
\end{center}

\subsection{Lossless case}\label{ssec:losless}
\begin{figure*}[htb]
\includegraphics[width=0.70\textwidth]{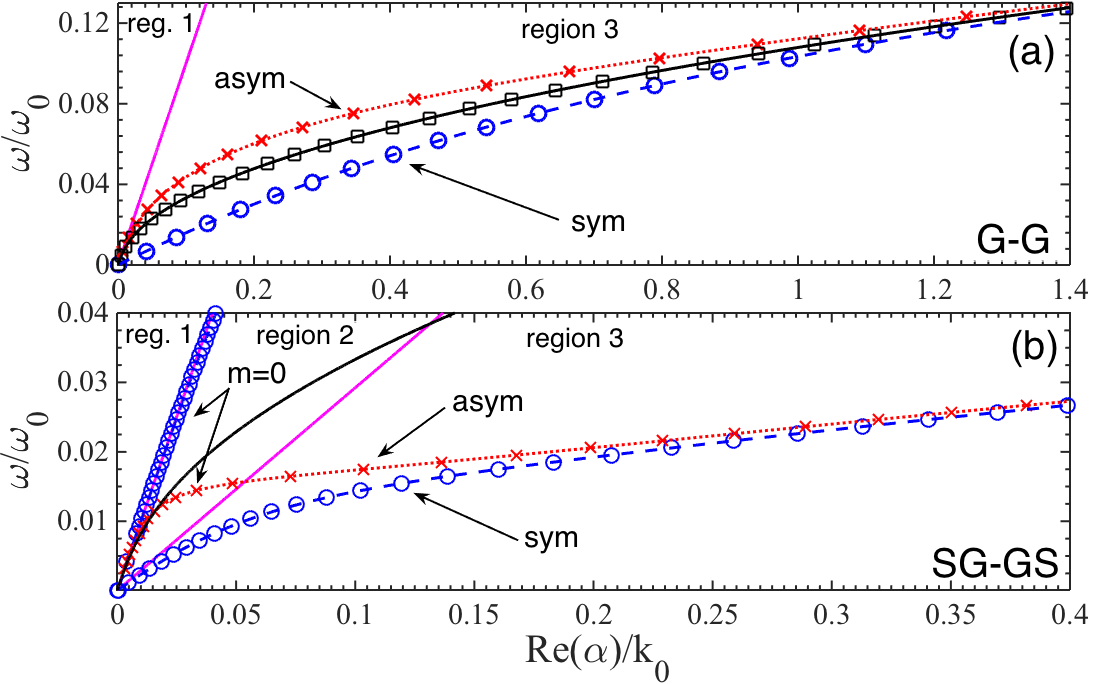}
\caption{\label{fig:dispertion}\footnotesize (color online). TM dispersion (real part) for $2a=0.4\mu$m,  $\mu_\textrm{F}=1$eV, $T=300$K,  $\Gamma=10^{11}$rad/s (identical figure obtained for $\Gamma=5\;10^{12}$rad/s), with $\lambda_0=1\mu$m, $\omega_0=2\pi c/\lambda_0$, $k_0=\omega_0/c$. Panel (a): G-G. Panel (b): SG-GS. with Si slabs of thickness $s=1\mu$m. Lines are calculated with the complete graphene conductivity, with symbols are calculated with the lossless approximation $\sigma=i \sigma_\textrm{I}$ and $\varepsilon=\varepsilon_{\textrm{R}}$. Solid black line corresponds to a single-graphene sheet TM dispersion $\al=k[1-4/(\sigma Z_0)^2]^{1/2}$. Purple  lines are the light cones $\omega=\al/c$ and $\omega=\al c/\sqrt{\varepsilon_{\textrm{R}}}$. }  
\end{figure*}

Now we discuss the particular case where the effects of losses are negligible in the structure, such that the slab dielectric permittivity is purely real $\varepsilon=\varepsilon_\textrm{R}$ and the graphene conductivity is purely imaginary $\sigma=i\sigma_\textrm{I}$.  This situation, which largely simplifies the discussion, can be fulfilled in practice: for instance at $\lambda=5\mu$m one has that $\varepsilon_\textrm{I}/\varepsilon_\textrm{R}< 10^{-5}$ for Silicon (Si) and $\sigma_\textrm{R}/\sigma_\textrm{I}< 10^{-3}$ for graphene [as we will see in section \ref{sec:SlaGra}]. Furthermore, for graphene  $\sigma_\textrm{R}/\sigma_\textrm{I}\ll1$ is realized below the graphene transition frequency $\omega_c=2 \mu_\textrm{F}/\hbar$ ($\mu_\textrm{F}$ being the chemical potential of the sheet, or equivalently its Fermi level), hence implying  $\sigma_\textrm{I}>0$. Under these assumptions, the propagation constant $\al$ is purely real, Eqs.~\eqref{eq:dispTE}-\eqref{eq:dispTM} can be recast in much simpler forms, and we can identify three regions on the $(\al,\omega)$ plane (cf. Fig. \ref{fig:dispertion}): 
\begin{itemize}
\item (i) region 1: $0\leq\al(\omega)\leq k$. It is on the left of the first light-cone, hence $\g{1}$ is purely imaginary while $\g{2}$ is real; 
\item (ii) region 2: $k<\al(\omega)< k\sqrt{\varepsilon}$. It corresponds to the area between the two light-cones, hence both $\g{i}$ are reals;
\item (iii) region 3: $\al(\omega)\geq k\sqrt{\varepsilon}$. It is on the right of the second light-cone, hence $\g{1}$ is real and $\g{2}$ is purely imaginary.
\end{itemize}
In the rest of this section we will derive the dispersion relation in the different regions, and summarize the results in table \ref{Tavola}.

\subsubsection{TE modes: Lossless case}\label{ssec:TEsDC}
In the lossless case, in region 1 equation (\ref{eq:dispTE}) has no guided waves solutions. In region 2, which is meaningful only in presence of the slabs ($\varepsilon\neq1$ and $s>0$), by isolating the term $\phi^2$ on one side of equation (\ref{eq:dispTE}), and imposing that the two sides should have the same phase (they have the same modulus, equal to 1) we obtain that the modes are the solutions $\al$ of the real equation
\begin{equation}\label{eq:TEreg2}
\g{2}s=\arctan\left(q\right)+\arctan\left(qF(\g{1}a)-\frac{i\;\et}{\g{2}}\right)+m\pi, 
\end{equation}
where different modes are labelled by natural numbers $m=0,1,2,3,\cdots$, and we introduced the real quantity $q=\g{1}/\g{2}$. In the absence of graphene, $\eta=0$, equation (\ref{eq:TEreg2}) reduces to the result of the slab-slab configuration  \cite{Riboli2008}. In this case, the symmetric mode dispersion function is below the antisymmetric one, both are continuous functions, and for $m=0$, the antisymmetric one has a non-zero lower frequency bound at $\omega^\textrm{Asym}_\textrm{cut-off}>0$ contrarily to the symmetric one which has $\omega^\textrm{Sym}_\textrm{cut-off}=0$.  The introduction of graphene in the structure ($\eta\neq0$) changes the dispersion functions, which tend to be globally shifted upwards in frequency, and now the $m=0$ symmetric mode dispersion function acquires a non-zero lower frequency bound $\omega^\textrm{Sym}_\textrm{cut-off}>0$.  Finally, it is worth stressing that Eq.  (\ref{eq:TEreg2}) can also be recast under the form 
\begin{equation}\label{eq:TEreg2form2}
\tan(\g{2}s)=\frac{q\left[1+F(\g{1}a)\right]-\frac{i\;\et}{\g{2}}}{1-q\left[qF(\g{1}a)-\frac{i\;\et}{\g{2}}\right]},
\end{equation}
which will be useful in deriving the expression of the LI pressure (see Section \ref{sec:press}).

In region 3, by introducing the real quantity $u=\g{1}/(-i\g{2})$ we can rewrite Eq.~\eqref{eq:dispTE} in the dimensionless real form
\begin{multline}\label{eq:dispTE3}
\phi\;(1-u)\left[u\;F(\g{1}a)+\frac{\et}{\g{2}}-1\right]+\\
\phi^{-1}\;(1+u)\left[u\;F(\g{1}a)+\frac{\et}{\g{2}}+1\right]=0.
\end{multline}
Now we can distinguish a first situation, corresponding to graphene-graphene configuration in absence of slabs. In this case $\varepsilon=1$, $\g{2}=i\g{1}$, $\g{1}>0$, $u=1$ hence  Eq.~\eqref{eq:dispTE3} becomes $\g{1}+\g{1}F(\g{1}a)+\sigma_\textrm{I} k Z_0=0$  which has no solution since we assumed $\sigma_\textrm{I}>0$. 
The remaining case is the graphene-graphene configuration in presence of slabs, so that $u\neq1$, for which it is easy to show that $g_\pm=\left[u\;F(\g{1}a)+\et/\g{2}\pm 1\right]\neq0$, and dividing Eq.~\eqref{eq:dispTE3} by $g_+$ one obtains the equation
\begin{equation}\label{eq:TEu3}
 \tanh(-i\g{2}s)=-\frac{u\;\left[1+F(\g{1}a)\right]+\frac{\eta}{\g{2}}}{1+u\;\left[u\;F(\g{1}a)+\frac{\eta}{\g{2}}\right]}, 
 \end{equation}
which clearly has no solutions (the two sides having opposite sign). In conclusion, in the lossless case, TE modes exist only in region 2 (hence in presence of the supporting slabs).

It is worth noticing that Eq. (\ref{eq:TEreg2form2}), which has been derived for $\g{2}$ purely real (region 2), reduces exactly to Eq. (\ref{eq:TEu3}) if one takes $\g{2}$ as purely imaginary.  And vice-versa, Eq. (\ref{eq:TEu3}), which has been derived for $\g{2}$ purely imaginary (region 3), reduces exactly to Eq. (\ref{eq:TEreg2form2})  if one takes $\g{2}$ as purely real.

\subsubsection{TM modes, lossless case}\label{ssec:TMsDC} 
Let us discuss, under the same lossless assumptions used in section \ref{ssec:TEsDC} for the TE modes, the presence of TM modes in the three regions of the $(\al,\omega)$ plane. In region 1, by definition $\g{1}$ is purely imaginary, hence, as for the TE case, no guided waves solutions are present. In region 2, following the same procedure as for the TE case, Eq. (\ref{eq:dispTM}) becomes the real equation:
\begin{equation}\label{eq:TMreg2}
\g{2}s=\arctan\left(p\right)+\arctan\left(\frac{p\;F(\g{1}a)}{1+i\de\;p\;F(\g{1}a)}\right)+m\pi, 
\end{equation}
where different modes are labelled by natural numbers $m=0,1,2,3,\cdots$, and where we introduced the real quantity $p=\g{1}/\gp{2}$. In the absence of graphene, $\de=0$, equation (\ref{eq:TMreg2}) reduces to the result of the slab-slab configuration  \cite{Riboli2008}. 

It is worth stressing that for $\de=0$, the symmetric mode dispersion function is below the antisymmetric one, both are continuous functions, and for $m=0$, and the antisymmetric one has a non-zero lower frequency bound at $\omega^\textrm{Asym}_\textrm{cut-off}>0$ contrarily to the symmetric one which has $\omega^\textrm{Sym}_\textrm{cut-off}=0$.  The introduction of graphene in the structure ($\de\neq0$) changes the dispersion functions, which in general, for $m>0$ tend to be globally shifted upwards in frequency. Remarkable is the case of the $m=0$ modes. Indeed, in presence of graphene the antisymmetric $m=0$ function splits into two branches: the lower branch is in the frequency region $(0\div\omega^\textrm{Asym}_{\textrm{cut-off},1})$, implying a zero-frequency lower frequency bound and with an upper bound; the upper branch is in the frequency region $(\omega^\textrm{Asym}_{\textrm{cut-off},2}\div\infty)$, with $\omega^\textrm{Asym}_{\textrm{cut-off},1}<\omega^\textrm{Asym}_{\textrm{cut-off},2}$. Between these two branches there is the $m=0$ symmetric dispersion function, which maintains a  zero frequency lower bound. It is worth noticing that the lowest of the two $m=0$ antisymmetric branches continues in region 3, perfectly matching the antisymmetric mode given by equation (\ref{eq:dispTM3}).

Finally, it is worth stressing that Eq.  (\ref{eq:TMreg2}) can also be recast under the form 
\begin{equation}\label{eq:TMreg2form2}
\tan(\g{2}s)=\frac{p\left[1+(1+i\de\;p)F(\g{1}a)\right]}{1-p(p-i\de)F(\g{1}a)},
\end{equation}
which will be useful in deriving the expression of the LI pressure.

In region 3, by introducing the real quantity $v=\g{1}/(-i\gp{2})$ we can rewrite Eq. (\ref{eq:dispTM}) in the dimensionless real form
\begin{multline}\label{eq:dispTM3}
\phi\;(v-1)\left[1-v\;(1-\de)\;F(\g{1}a)\right]+\\
\phi^{-1}\;(v+1)\left[1+v\;(1+\de)\;F(\g{1}a)\right]=0.
\end{multline}
Now we can distinguish a first case, corresponding to graphene-graphene configuration in absence of slabs. In this case $\varepsilon=1$, $\g{2}=i\g{1}$, $\g{1}>0$, $v=1$, then  Eq. (\ref{eq:dispTM3}) becomes
\begin{equation}\label{eq:ggr3TM}
1+(1+\de)\;F(\g{1}a)=0,
\end{equation}
which admits (both symmetric and antisymmetric mode) solutions, contrarily to the corresponding TE case. 
It is worth investigating the limit $a\rightarrow0$ of Eq.~\eqref{eq:ggr3TM}, for which it is easy to show that the propagation constant for the symmetric mode diverges as $a^{-1/2}$, while it is finite for the  antisymmetric case: 
\begin{align} 
\al^{\textrm{s}}_{0}&\sim\sqrt{\frac{k}{\sigma_\textrm{I} Z_0}}\;\;\frac{1}{a^{1/2}},\label{eq:allimsym}\\
\al^{\textrm{a}}_{0}& = k\sqrt{1+\frac{1}{\sigma_\textrm{I}^2 Z_0^2}}+\frac{k^2}{\sigma_\textrm{I}^2 Z_0^2\sqrt{1+\sigma_\textrm{I}^2 Z_0^2}}\;a.\label{eq:allimasym}
\end{align}
The lack of a finite value for the symmetric mode propagation constant in this limit is in accordance with the fact that a single graphene sheet supports only the antisymmetric mode ($H_y$ is antisymmetric, it exhibits a jump at the interface, and its dispersion relation is $\sigma_\textrm{I}Z_0\g{1}=2k$). The fact that $\al^{\textrm{s}}$ can reach very large values at small separations will be a crucial feature in the investigation of the LI force. This effect will remain valid also in presence of supporting slabs. In figure \ref{fig:pressures} panels (m-n-o) we plot $\al(\omega,a)$ as a function of the separation $2a$, and such asymptotic behaviors can be recognized.

The remaining case is the graphene-graphene configuration in presence of supporting slabs, so that $v\neq1$. In this case, following a procedure similar to that used for the TE case [see Eq. (\ref{eq:TEu3})], we obtain that Eq. (\ref{eq:dispTM3}) can be written as 
\begin{equation}\label{eq:TMu3}
 \tanh(-i\g{2}s)=-\frac{v\;\left[1+(1+\de\;v)\;F(\g{1}a)\right]}{1+v\;(v+\de)\;F(\g{1}a)}. 
 \end{equation}
This equation has solutions provided that the graphene is present. Indeed for the simple slab-slab configuration ($\de=0$) the two sides of Eq. (\ref{eq:TMu3}) have opposite sign.

It is worth noticing that,  analogously to the TE case,  Eq.  (\ref{eq:TMreg2form2}), which has been derived for $\g{2}$ purely real (region 2), reduces exactly to Eq. (\ref{eq:TMu3}) if one takes $\g{2}$ as purely imaginary.  And vice-versa, Eq. (\ref{eq:TMu3}), which has been derived for $\g{2}$ purely imaginary (region 3), reduces exactly to Eq. (\ref{eq:TMreg2form2})  if one takes $\g{2}$ as purely real. This means that in both regions 2 and 3 one can use only Eq. (\ref{eq:TMreg2form2}) [or only Eq. (\ref{eq:TMu3})]. Such a property will allow to derive a unique expression for the TM LI pressure valid in both regions (see Section \ref{sec:press}).

Figure \ref{fig:dispertion} shows that the s/a TM dispersions relations within the lossless approximation (symbols) reproduce perfectly the lossy model results, for both the G-G and SG-GS configurations. It is worth noticing that the G-G dispersion relation (entirely in region 3) increases very slowly and reaches, at a given frequency, values of $\al$ larger than those of a dielectric waveguide. Figure \ref{fig:dispertion} also shows that the effect of introducing supporting slabs is to add modes in region 2, and to bend even further the dispersion curve.

\section{Slabs and  Graphene Sheets Optical Properties}\label{sec:SlaGra}
We will consider slabs made of Silicon (Si) with dielectric permittivity $\varepsilon(\omega)=\varepsilon_\textrm{R}(\omega)+i \varepsilon_\textrm{I}(\omega)$ extracted from the Palik's handbook  \cite{Palik}. The graphene conductivity $\sigma(\omega)=\sigma_\textrm{R}(\omega)+i\sigma_\textrm{I}(\omega)$, for a gapless doped graphene sheet, is modeled as the sum of the intra-band (Drude like) and inter-band contributions  \cite{Falkovsky2008,Abajo2011,Ferrari2015}:
\begin{eqnarray}
\sigma(\omega)&=&\sigma_\textrm{intra}(\omega)+\sigma_\textrm{inter}(\omega), \label{eq:sigma}\\
\sigma_\textrm{intra}(\omega)&=&\frac{i8\sigma_0\;k_BT}{\pi(\hbar\omega+i\hbar\Gamma)}\ln\left[2\;\cosh\left(\frac{\mu_\textrm{F}}{2k_BT}\right)\right], \nonumber\\
\sigma_\textrm{inter}(\omega)&=&
\sigma_0\left[\mathcal{G}\left(\frac{\hbar\omega}{2}\right)+i\frac{4\hbar\omega}{\pi}\int_0^{\infty}\frac{\mathcal{G}\left(\xi\right)-\mathcal{G}\left(\frac{\hbar\omega}{2}\right)}{(\hbar\omega)^2-4\xi^2}d\xi\right],\nonumber
\end{eqnarray}
where $\sigma_0=e^2/(4\hbar)$, $e$ is the electron charge, $T$ is the temperature of the sheet, $\mu_\textrm{F}$ is the chemical potential (or equivalently the Fermi level), and $\mathcal{G}(x)=\sinh(x/k_BT)/[\cosh(\mu_\textrm{F}/k_BT)+\cosh(x/k_BT)]$. The quantity $\Gamma=1/\tau$ is the inverse of the relaxation time, and depends on the electronic collision mechanisms. 
One of the most interesting properties of graphene is the possibility to tune its conductivity by changing its chemical potential $\mu_\textrm{F}$ (typically $0\div1$eV), and this can be done via chemical doping or electrostatic doping realized by simply applying a voltage to the sheet. 

\begin{figure}[htb]
\includegraphics[width=0.48\textwidth]{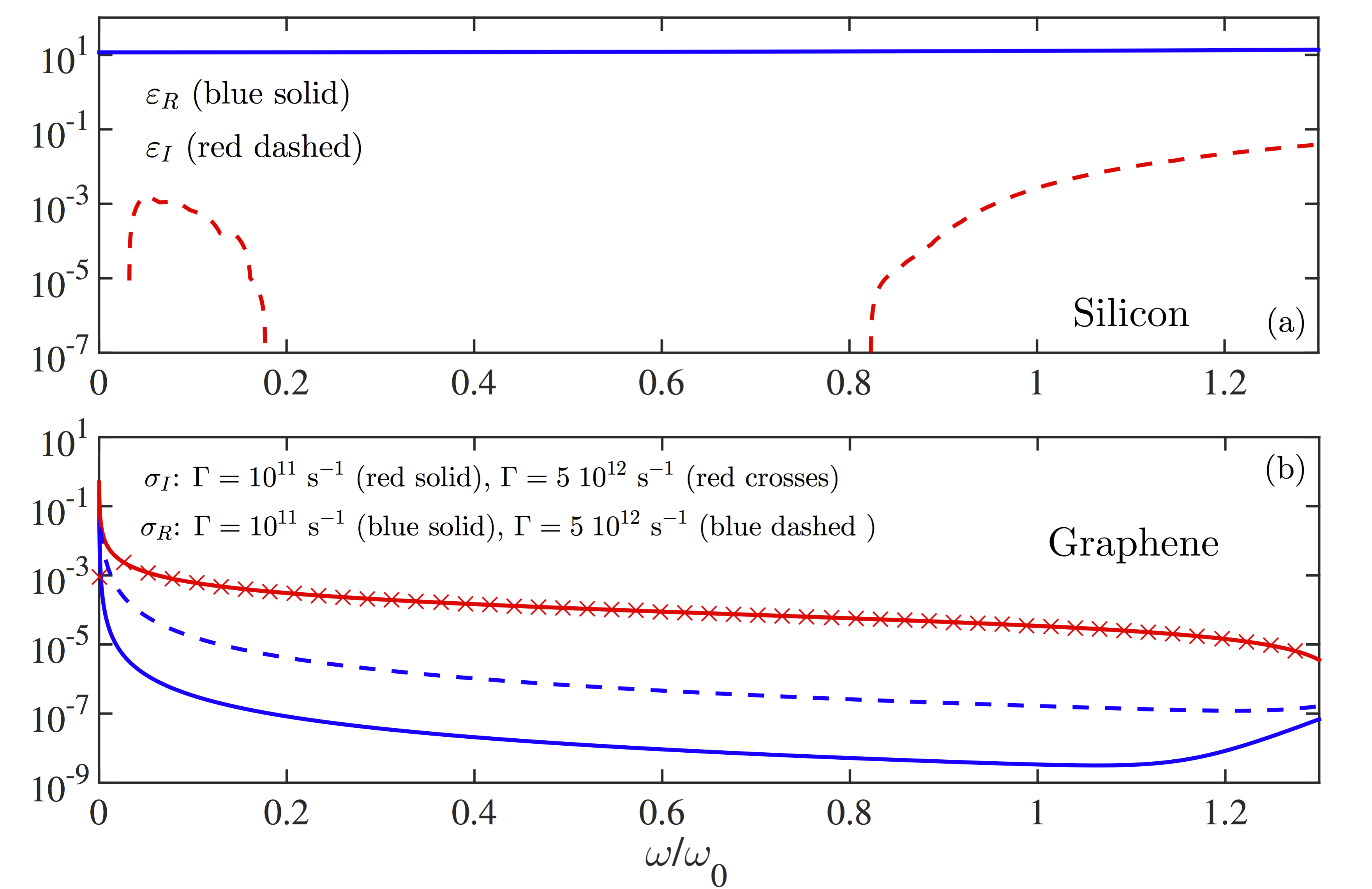}
\caption{\label{fig:optical_functions}\footnotesize (color online). Optical properties of the slabs and graphene sheets. Frequencies are in units of $\omega_0=2\pi c/\lambda_0$, $\lambda_0=1\mu$m. Panel (a): real (blue, solid line) and imaginary (red, dashed line) parts of the Silicon dielectric permittivity \cite{Palik} $\varepsilon(\omega)$. Panel (b): Graphene conductivity $\sigma(\omega)$,  with $\mu_\textrm{F}=1$eV, and $T=300$K. Real part (blue solid line for $\Gamma=10^{11}$rad/s and blue dashed line for $\Gamma=5\;10^{12}$rad/s) and imaginary part (red solid line for $\Gamma=10^{11}$rad/s and red crosses for $\Gamma=5\;10^{12}$rad/s).}
\end{figure}

In Figure \ref{fig:optical_functions} we plot the Si dielectric permittivity and the graphene conductivity. The figure  
shows the presence of a wide region where both ratios $\varepsilon_\textrm{I}/\varepsilon_\textrm{R}$ and $\sigma_\textrm{R}/\sigma_\textrm{I}$  (and hence losses)  are considerably small.
It is worth stressing that, in order to remain in such a lossless condition for graphene,  $\omega$ should: (i) not be too small (in the limit of small frequencies $\sigma_\textrm{R}>0$ while $\sigma_\textrm{I}=0$); but also (ii) be much smaller than the graphene transition which takes place at $\omega_{\textrm{c}}=2\mu_\textrm{F}/\hbar$, indeed close to such a value $\sigma_\textrm{R}>0$ and $\sigma_\textrm{I}<0$. By decreasing the value of $\mu_\textrm{F}$, the frequency range where graphene can be considered lossless becomes smaller and smaller. Hence the condition $0\ll\omega\ll2\mu_\textrm{F}$ must be fulfilled. In Figure \ref{fig:optical_functions}(b) we used $\mu_\textrm{F}=1$eV$=0.81 \hbar\omega_0$ (with $\omega_0=2\pi c/\lambda_0$, $\lambda_0=1\mu$m). This corresponds to a transition at $\omega_{\textrm{c}}/\omega_0=2\mu_\textrm{F}/(\hbar \omega_0)=1.62$, and indeed at frequencies $\omega/\omega_0\approx1.2$ we start seeing a clear change in the conductivity which delimitates the lossless frequency range.
In practice, in the calculations of the radiation pressure in section \ref{sec:numres}, we will use $\omega/\omega_0=0.2$ (i.e. $\lambda=5\mu$m), $\omega/\omega_0=0.645$ (i.e. $\lambda=1.55\mu$m), and $\omega/\omega_0=0.125$ (i.e. $\lambda=8\mu$m). By analyzing  the graphene conductivity function we see that this requires to set the Fermi level $\mu_\textrm{F}>0.3$eV, $\mu_\textrm{F}>0.6$eV, $\mu_\textrm{F}>0.7$eV respectively, in order to fulfill the lossless condition.

\begin{figure}[htb]
\includegraphics[width=0.48\textwidth]{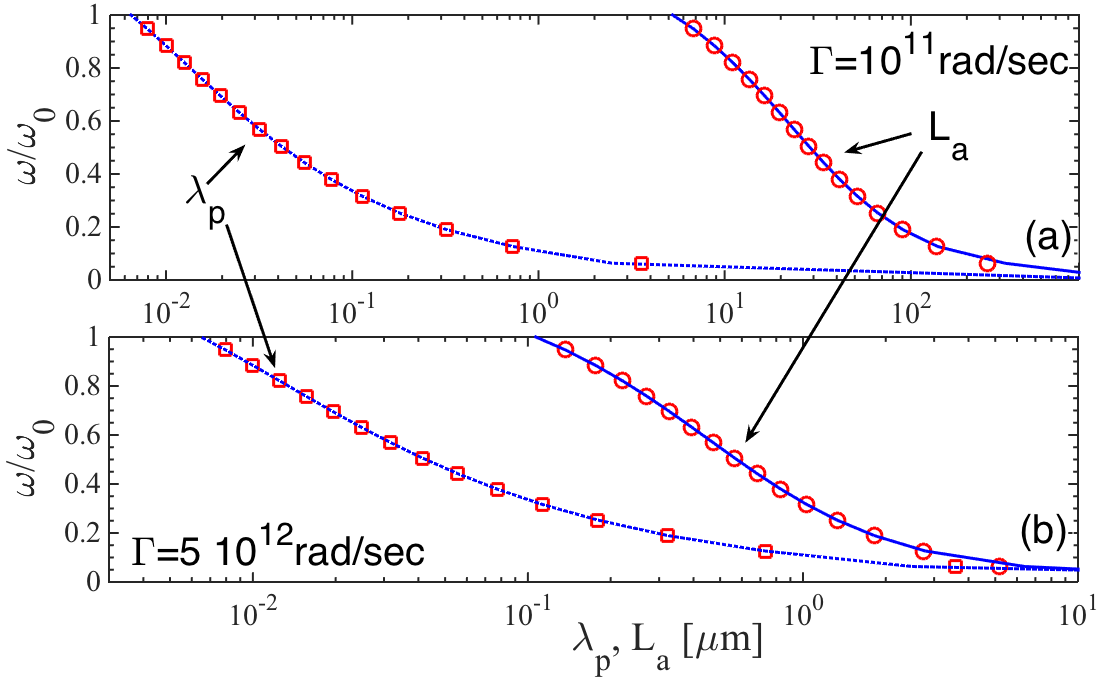}
\caption{\label{fig:lengths}\footnotesize (color online). Length scales $\lambda_\textrm{p}$ (dotted-blue line for symmetric modes, red squares for antisymmetric ones), and $L_\textrm{a}$ (solid-blue line for symmetric modes, red circles for antisymmetric ones), for the TM mode, G-G with $2a=0.4\mu$m, $\mu_\textrm{F}=1$eV. (a): $\Gamma=10^{11}$rad/s. (b): $\Gamma=5\;10^{12}$rad/s.}
\end{figure}

\section{Length scales}\label{sec:Length} 
To fulfill the conditions of validity of Eq.~\eqref{eq:LIpress} for the LI pressure (i.e. lossless case and infinitely extended waveguides in the $xy$ plane), it is necessary to investigate the length-scales associated to the excited light mode $\al(\omega)$: (i) the mode propagation wavelength $\lambda_\textrm{p}=2\pi/\textrm{Re}(\al)$, and (ii) the mode absorption length $L_\textrm{a}=1/[2\textrm{Im}(\al)]$,  characterizing the wave intensity decay.  In order to minimize the boundary effects due to the finite extension of the system in the $x$ direction, the waveguide length $L_x$ must be much larger than $\lambda_\textrm{p}$, such that the wave possesses several oscillations at the scale of the system length. Furthermore, in order to assume that the intensity of the wave is as much constant as possible in the $x$ direction, the absorption length must be much larger than $L_x$. In practice we need to satisfy for $L_x$ the \emph{length condition}: 
\begin{equation}\label{eq:lc}
\lambda_\textrm{p}\ll L_x\ll L_\textrm{a}.
\end{equation}
This implies finding a configuration where $\textrm{Re}(\al)\gg\textrm{Im}(\al)$, i.e. a system as lossless as possible. 

Figure \ref{fig:lengths} shows, for the TM modes of G-G, both $\lambda_\textrm{p}$ and $L_\textrm{a}$  (see caption for details) for two different values of $\Gamma$. We see that it is possible to find a vast region (stuck between the two curves) where the length condition is satisfied, and that the effect of losses is to reduce such a region.

\section{Light-Induced Pressure}\label{sec:press}
The LI pressure linearly depends on the intensity of light in the structure, so that it is useful to introduce the power linear density per unit of length $W_y$ in the direction of invariance $y$, for a given mode:
\begin{multline}\label{eq:power}
\mathcal{P}=\frac{1}{W_y}\int_{0}^{W_y}dy \int_{-\infty}^{\infty}     \langle S_x  \rangle_t dz
 =\\
 \frac{1}{2}\int_{-\infty}^{\infty} \textrm{Re}\left[  E_yH_z^*-E_zH_y^* \right]  dz,
\end{multline}  
where ${\bf S}$ is the Poynting vector. 
It can be shown that $p_{\textrm{LI}}$ is proportional to a coefficient depending on the field amplitudes. In order to find a closed form expression for  $p_{\textrm{LI}}$ we can derive $\mathcal{P}$ in terms of the same coefficient appearing in $p_{\textrm{LI}}$. Hence, after eliminating the common coefficient  \cite{Riboli2008}, we can express $p_{\textrm{LI}}$ in terms of $\mathcal{P}$.

Let us start by considering the s/a LI pressure for the G-G configuration, which in the lossless approximation (see table \ref{Tavola}) can exists only in region 3 and for the TM mode.  By using Eq.~\eqref{eq:LIpress}, and the dispersion relation, it can be explicitly calculated providing the expression:
\begin{equation} 
p_{\textrm{TM}}^{\textrm{s/a}}=\frac{\mathcal{P}_{\textrm{TM}}^{\textrm{s/a}}}{2 \omega \alpha_{\textrm{TM}}^{\textrm{s/a}}}\;\;\frac{\g{1}^3 (\de+2)}{1-a\g{1}(\de+2)}. \label{eq:pressure_TM_gra}
\end{equation}
It is wort investigating the limit of this expression for $a\rightarrow 0$. By using Eqs.\eqref{eq:allimsym} and \eqref{eq:allimasym} we obtain that pressure (\ref{eq:pressure_TM_gra}) diverges as $-a^{-3/2}$ for the symmetric mode, while it is positive and finite for the antisymmetric one:
\begin{align} 
p_{\textrm{TM},0}^{\textrm{s}}&\sim-\frac{\mathcal{P}_{\textrm{TM}}^{\textrm{s}}}{4c}\;\;\frac{1}{\sqrt{k \sigma_\textrm{I} Z_0}}\;\;\frac{1}{a^{3/2}},\label{eq:pressyma0}\\
p_{\textrm{TM},0}^{\textrm{a}}&=\frac{\mathcal{P}_{\textrm{TM}}^{\textrm{a}}}{2c }\;\;\frac{k}{\sigma_\textrm{I}^2 Z_0^2\sqrt{1+\sigma_\textrm{I}^2 Z_0^2}}.\label{eq:presasyma0}
\end{align}
Following the same procedure used for G-G, we derive the TE/TM s/a modes LI pressure for the SG-GS configuration:
\begin{widetext}
\begin{align}
p_{\textrm{TE}}^{\textrm{s/a}}&=-\frac{\mathcal{P}_{\textrm{TE}}^{\textrm{s/a}}G(\g{1}a)\g{1}^2}{2\omega\alpha_{\textrm{TE}}^{\textrm{s/a}}}\left\{ \left[1+ \left(q F(\g{1}a)-\frac{i\eta}{\g{2}}\right)^2\right] \left(s+\frac{1}{\g{1}}\right) + a  G(\g{1}a) +\frac{(1+q^2)F(\g{1}a)}{\g{1}}-\frac{i\eta q}{\g{2}\g{1}} \right\}^{-1}, \label{eq:pressure_TE}\\
p_{\textrm{TM}}^{\textrm{s/a}}&=-\frac{\mathcal{P}_{\textrm{TM}}^{\textrm{s/a}}G(\g{1}a)\varepsilon^2\g{1}^2}{2\omega\alpha_{\textrm{TM}}^{\textrm{s/a}}} 
\left\{ \left[1+ \left(p F(\g{1}a)+i\de\right)^2+\de^2\left(1-p^2 F(\g{1}a)^2\right)\right]\left(s\varepsilon+ \frac{1}{\g{1}}\frac{\varepsilon^2+p^2}{1+p^2}\right)+\;a\varepsilon^2G(\g{1}a)\right.\notag\\
&\left. \;\; \;\;\;\;\;\;\;\;\;\;\;\;\;\;\;\;\;\;\;\;\;\;\;\;\;\;\;\;\;\;\;\;\;\;\; \;\;\;\;\;\;\;\;\;\;\;\;\;\;\;\;\;\;\;\;\;\;\;\;\;\;\;\;\;\;\;\;\;\;\;\;\;\;\;\;\;\;\;\;\;\;\;\;\;\;\;\;\;\;\;\;\;\;\;\;\;\;\;\;\;\;+\frac{(\varepsilon^2+p^2)F(\g{1}a)}{\g{1}}+\frac{i\de p^3 F(\g{1}a)^2 }{\g{1}}\right\}^{-1}, \label{eq:pressure_TM}
\end{align}  
 \end{widetext}
with $G(x)=1-F(x)^2$, $k=\omega/c$, $q=\g{1}/\g{2}$, $p=\g{1}/\gp{2}$. Note that  Eq.~\eqref{eq:pressure_TE} is valid only in region 2 (in region 3 there are no TE modes), while Eq.~\eqref{eq:pressure_TM} is valid in both regions 2 and 3. It is worth noticing that in region 2, and in the absence of graphene ($\de=\eta=0$), Eqs.~\eqref{eq:pressure_TE} and \eqref{eq:pressure_TM} reproduce the slab-slab expressions derived in  \cite{Riboli2008}.

It is remarkable that, in the lossless case, the LI pressure $p_{\textrm{LI}}$ can be calculated without evaluating the Maxwell stress tensor  \cite{Povinelli2005,Rakich2009}: 
\begin{equation}\label{eq:simplpress_old}
p_{\textrm{LI}}=-\frac{\mathcal{P}}{2\omega}\;\frac{1}{v_g}\partial_a\omega(\al,a), 
\end{equation}
where  $\al(\omega,a)$ is the dispersion relation, and  $v_g=\partial_{\al}\omega(\al,a)$ is the group velocity.
By using the lossless condition $d\omega=0$, we have $\partial_a\omega\;da+\partial_{\al}\omega\;d\al=0$ and $d\al=\partial_a\al\;da$, which together give for the group velocity $v_g=\partial_{\al}\omega(\al,a)=-\partial_a\omega(\al,a)/\partial_a\al(\omega,a)$, and hence that Eq. (\ref{eq:simplpress_old}) can be recast as:
\begin{equation}\label{eq:simplpress}
p_{\textrm{LI}}=\frac{\mathcal{P}}{2\omega}\;\partial_a\al(\omega,a),
\end{equation}
where the pressure is expressed as a simple derivative of the dispersion relation with respect to the half separation distance $a$. Expression \eqref{eq:simplpress} permits an immediate derivation of   Eqs.~\eqref{eq:pressyma0}-\eqref{eq:presasyma0} using  Eqs.~\eqref{eq:allimsym}-\eqref{eq:allimasym}. For arbitrary separations, the derivative should be calculated numerically, hiding the explicit parameter dependences, which is instead present in expressions \eqref{eq:pressure_TM_gra}, \eqref{eq:pressure_TE}, \eqref{eq:pressure_TM}. In figure \ref{fig:pressures} panels (m=n=o) we plot $\al(\omega,a)$ as a function of the separation $2a$.

\section{Casimir-Lifshitz force}\label{sec:Casimir}
Even in the absence of additional excited modes, both vacuum ($T=0$) and thermal fluctuations of the electromagnetic field give rise to the so called Casimir-Lifshitz force, which becomes large at small separations between the objects. In this section we provide the expression of the CL pressure between systems containing graphene sheets  \cite{variCasimirGraphene}, and in particular the G-G and SG-GS configurations. 
\begin{figure}[htb]
\includegraphics[width=0.48\textwidth]{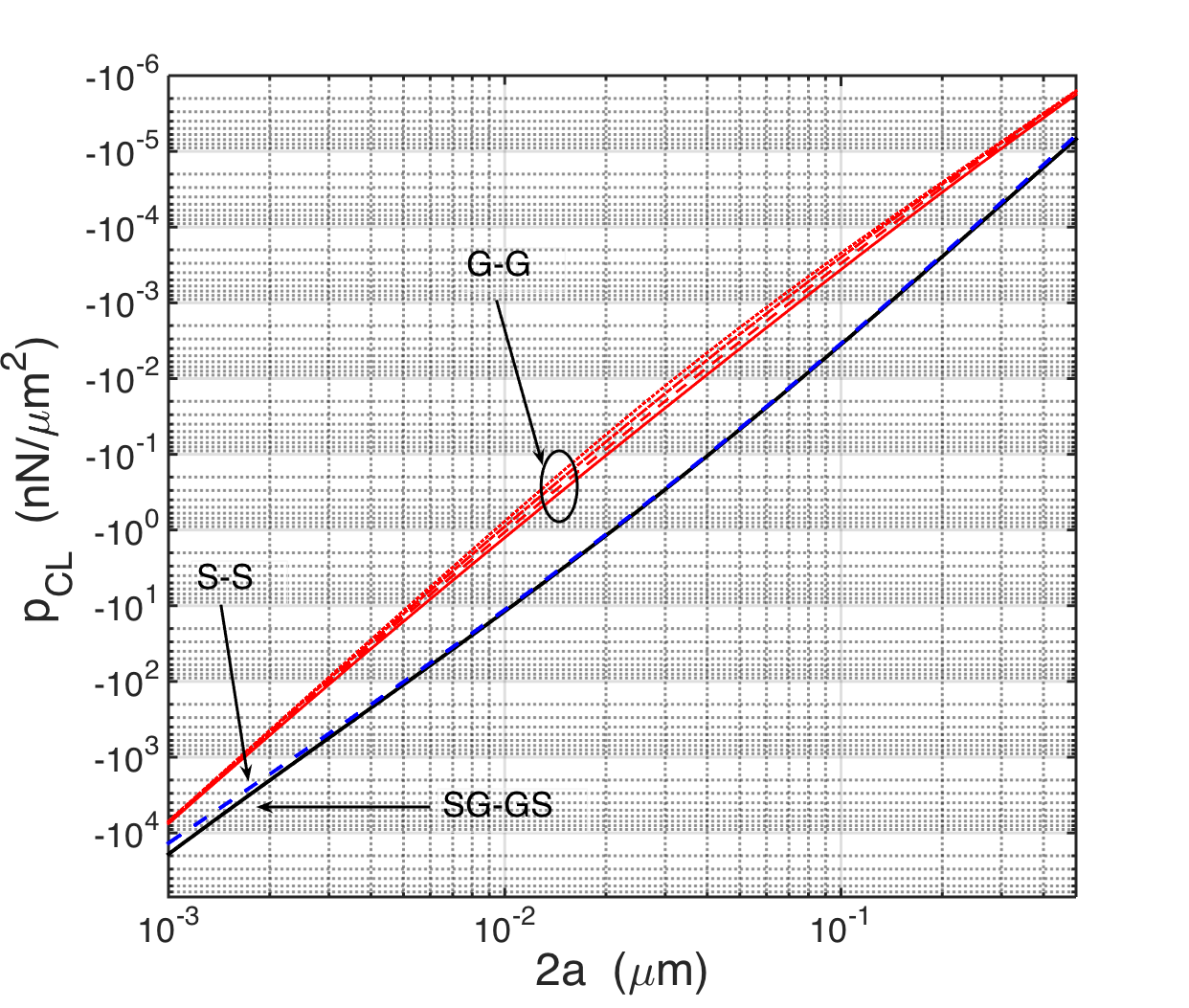}
\caption{\label{fig:fig_pCL}\footnotesize (color online). Casimir-Lifshitz pressure Eq.~\eqref{eq:Casimir} at $T=300K$. Dashed-blue line: Silicon Slab-Slab configuration (S-S) with slab thickness $s=1\mu$m. Solid dark line: SG-GS configuration, $s=1\mu$m, $\Gamma=5\;10^{12}$rad/sec, $\mu_{\textrm{F}}=0.3$eV, $0.6$eV, $1.0$eV (lines corresponding to the three values of $\mu_{\textrm{F}}$ are not distinguishable). Red lines: G-G configuration, with $\Gamma=5\;10^{12}$rad/sec,  $\mu_{\textrm{F}}=0$eV (dotted), $\mu_{\textrm{F}}=0.3$eV (dash-dotted),  $\mu_{\textrm{F}}=0.6$eV (dashed), $\mu_{\textrm{F}}=1.0$eV (solid).}
\end{figure}
The Casimir-Lifshitz interaction is the result of the sum over all modes of the field, which implies the integration over entire frequency and wave vector spaces. This means that  the complete complex permittivity and conductivity functions $\varepsilon$ and $\sigma$ are required. The Casimir-Lifshitz pressure is given by:
\begin{equation}\label{eq:Casimir}
p_{\textrm{CL}}=-\frac{k_BT}{\pi}  \sum_{n=0}^{\infty}\sideset{'}{}\int_0^{\infty}\textrm{d}QQq\sum_{p}\frac{1}{\rho_p^{-2}\;e^{2 q d}-1};
\end{equation}
where the prime ${\prime}$ on the sum means that the $n=0$ term should be multiplied by $1/2$. Here $d=2a$ is the separation between the two bodies, $p=\textrm{TE,TM}$ are the two polarizations, a rotation on the complex frequency plane has been performed, hence  \cite{LLelec} $\varepsilon=\varepsilon(i\;\xi_n)=1+\frac{2}{\pi}\int_0^\infty\frac{\omega\varepsilon_{\textrm{I}}(\omega)}{\omega^2+\xi_n^2}\textrm{d}\omega$  (see figure \ref{fig:fig_SiSig_rot}(a) for the case of Silicon), $\sigma=\sigma(i\;\xi_n)$ using just the analytical form (\ref{eq:sigma}) at imaginary frequencies [see figure \ref{fig:fig_SiSig_rot}(b), where different values of $\mu_{\textrm{F}}$ are considered], $\xi_n= 2\pi k_B T n/\hbar$, $q  = \sqrt{\xi_n^2+Q^2}$,  and $\rho_p$ are the reflection coefficients of the GS block, i.e. that of a wave impinging on a single graphene sheet sustained by a dielectric slab of thickness $s$. The reflection coefficients of the graphene-slab bilayer can be derived from Maxwell equations and boundary conditions analogous to those used in section \ref{sec:DispRel}:
\begin{align} 
\rho_{\textrm{TE}} &= \frac{(q_s+q)(q_s-q^-) - \phi^2(q_s-q)(q_s+q^-)}{\phi^2(q_s-q)(q_s-q^{+}) - (q_s+q)(q_s+q^{+})};\label{eq:rhoTE}\\
\rho_{\textrm{TM}} &= \frac{(q'_s+q)(q'_s-q-\beta) - \phi^2(q'_s-q)(q'_s+q-\beta)}{\phi^2(q'_s-q)(q'_s-q+\beta) - (q'_s+q)(q'_s+q+\beta)},\label{eq:rhoTM}
\end{align}
where $q_s  = \sqrt{\varepsilon\xi_n^2+Q^2}$, $q'_s   = q_s/\varepsilon$, , $\phi =e^{-q_s s}$,  $\beta= qq'_s\sigma Z_0/(\xi_n/c)$, $q^{\pm}  =  q\pm \sigma Z_0 (\xi_n/c)$.
\begin{figure}[htb]
\includegraphics[width=0.48\textwidth]{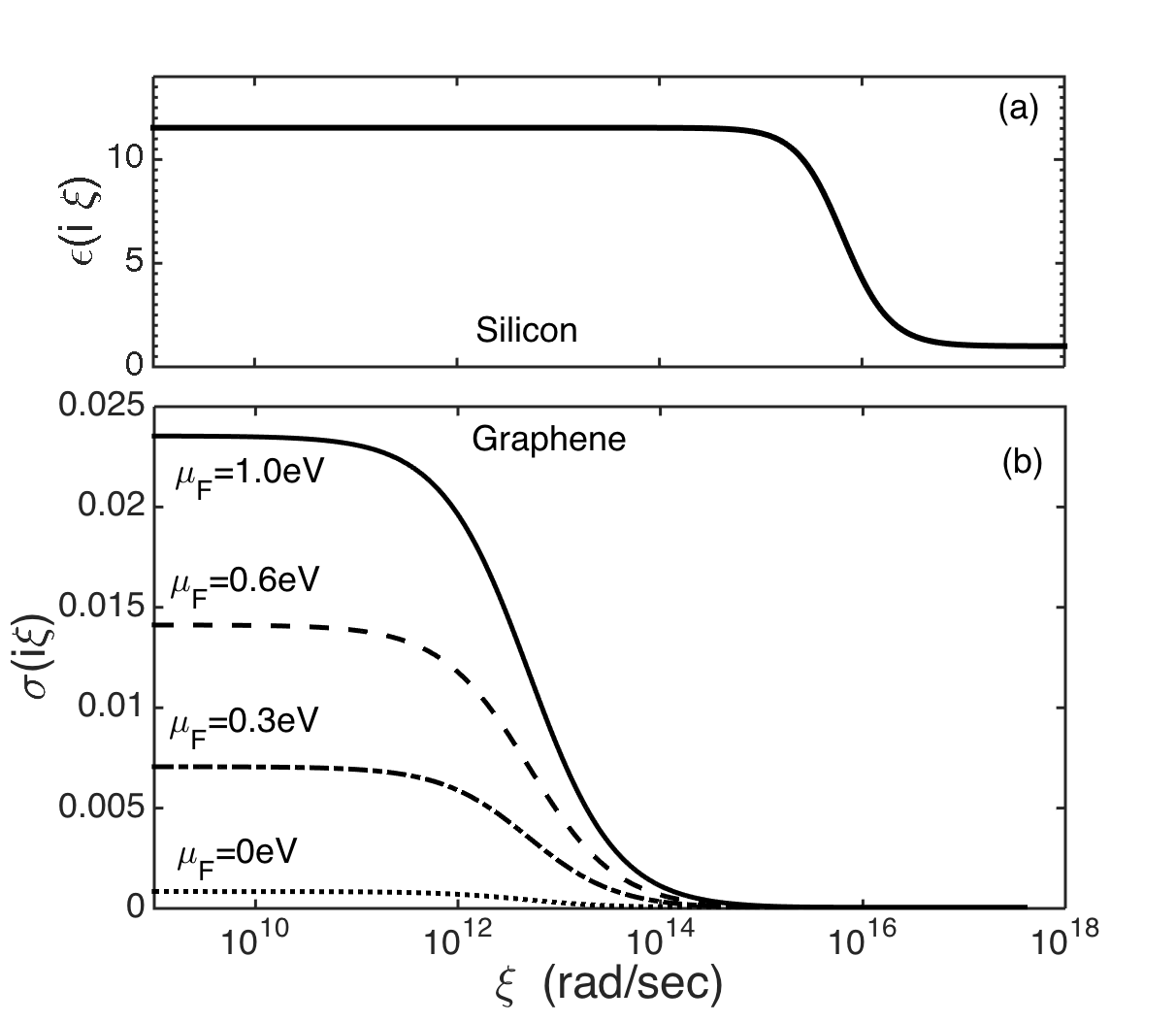}
\caption{\label{fig:fig_SiSig_rot}\footnotesize Panel (a): Silicon dielectric permittivity at complex frequencies $\varepsilon(i\xi)$ obtained by integration over experimental data of $\varepsilon_{\textrm{I}}(\omega)$ (see section \ref{sec:SlaGra}). Panel(b): Graphene conductivity at complex frequencies $\sigma(i\xi)$ obtained by equation (\ref{eq:sigma}), with $T=300K$, $\Gamma=5\;10^{12}$rad/sec, and $\mu_{\textrm{F}}=0$eV (dotted), $\mu_{\textrm{F}}=0.3$eV (dash-dotted),  $\mu_{\textrm{F}}=0.6$eV (dashed), $\mu_{\textrm{F}}=1.0$eV (solid).}
\end{figure}
They reproduce the single graphene sheet reflection coefficients by setting $s=0$ and $\varepsilon=1$ , and the single slab Fresnel reflection coefficients by setting $\beta=0$ and $q^{\pm}  =  q$. The limiting case of a dielectric occupying the entire half-space is obtained by setting $\phi=0$. 
In figure (\ref{fig:fig_pCL}) we plot the CL pressure for the G-G (red lines) and  SG-GS (blue lines) for different values of the graphene chemical potential. We also plot the CL between two slabs in absence of graphene (S-S).
The pressure is always attractive, at small separations scales as $1/a^{4}$ for G-G configuration with $\mu_{\textrm{F}}=0$eV, while as $1/a^{3}$ for the S-S configuration. We see that the CL pressure for G-G is much weaker than for SG-GS. The CL pressure for SG-GS is practically insensitive to the variation of the chemical potential (the 4 curves overlap), and coincides practically for all separations with the pressure of the S-S configuration. 
To give a idea about the number of frequencies $\xi_n$ used in the sum (\ref{eq:Casimir}): for smallest distance $2a=1nm$ we needed $n_{max}\approx 1200$.  Of course the calculated values for the CL pressure should be considered as an approximation at the extremely small separation of $2a=1nm$, where non-local effects for the graphene conductivity may possibly start playing a role.

\section{Numerical results and discussions}\label{sec:numres}
Using the expression derived in the previous sections, here we evaluate the LI and CL pressures for both G-G and SG-GS configurations, as a function of the waveguide separation $2a$, and of the chemical potential $\mu_\textrm{F}$.

\begin{figure*}[htb]
\begin{center}
\includegraphics[width=6.25cm,clip=]{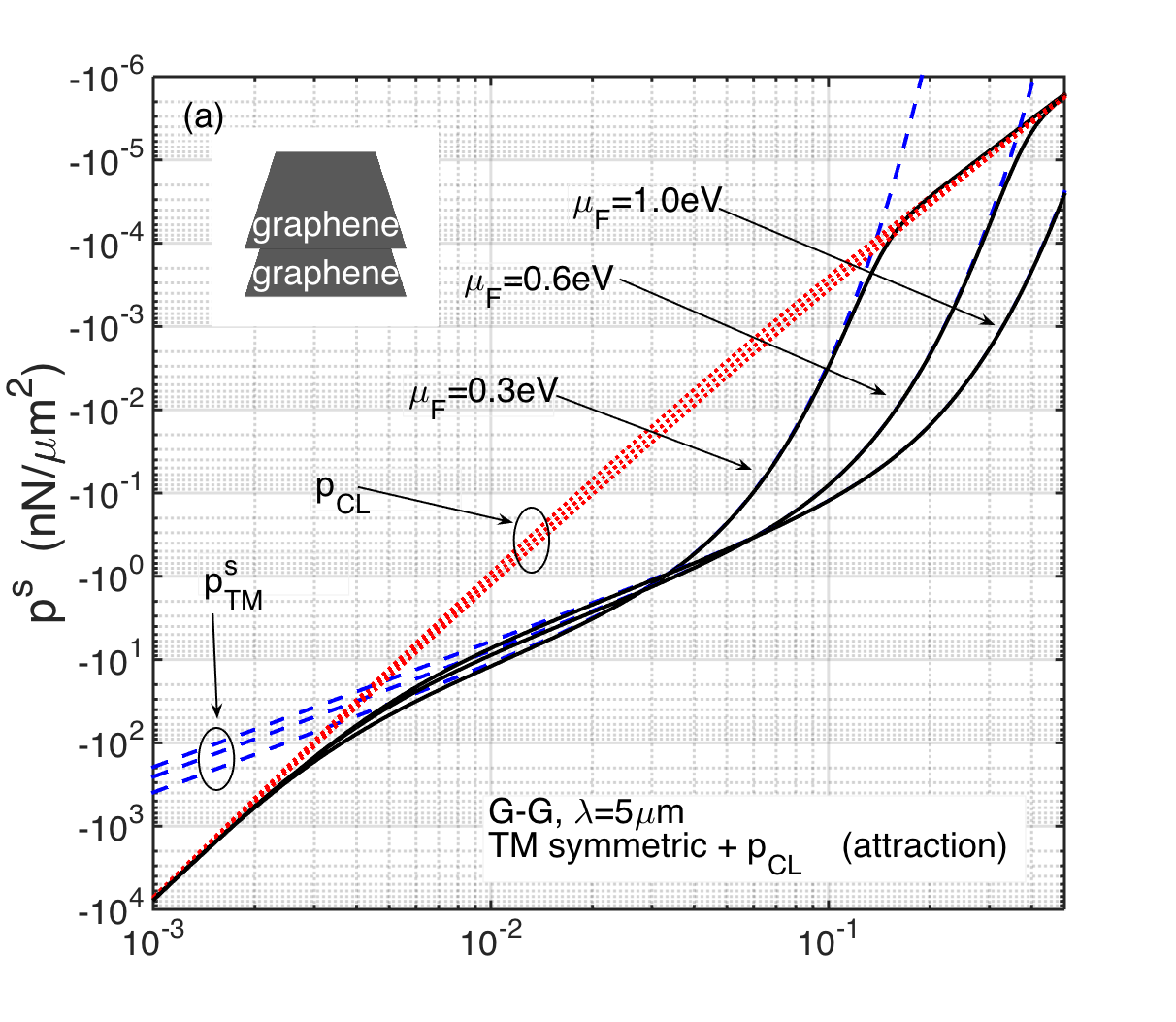}\hspace{-.43cm}
\includegraphics[width=6.25cm,clip=]{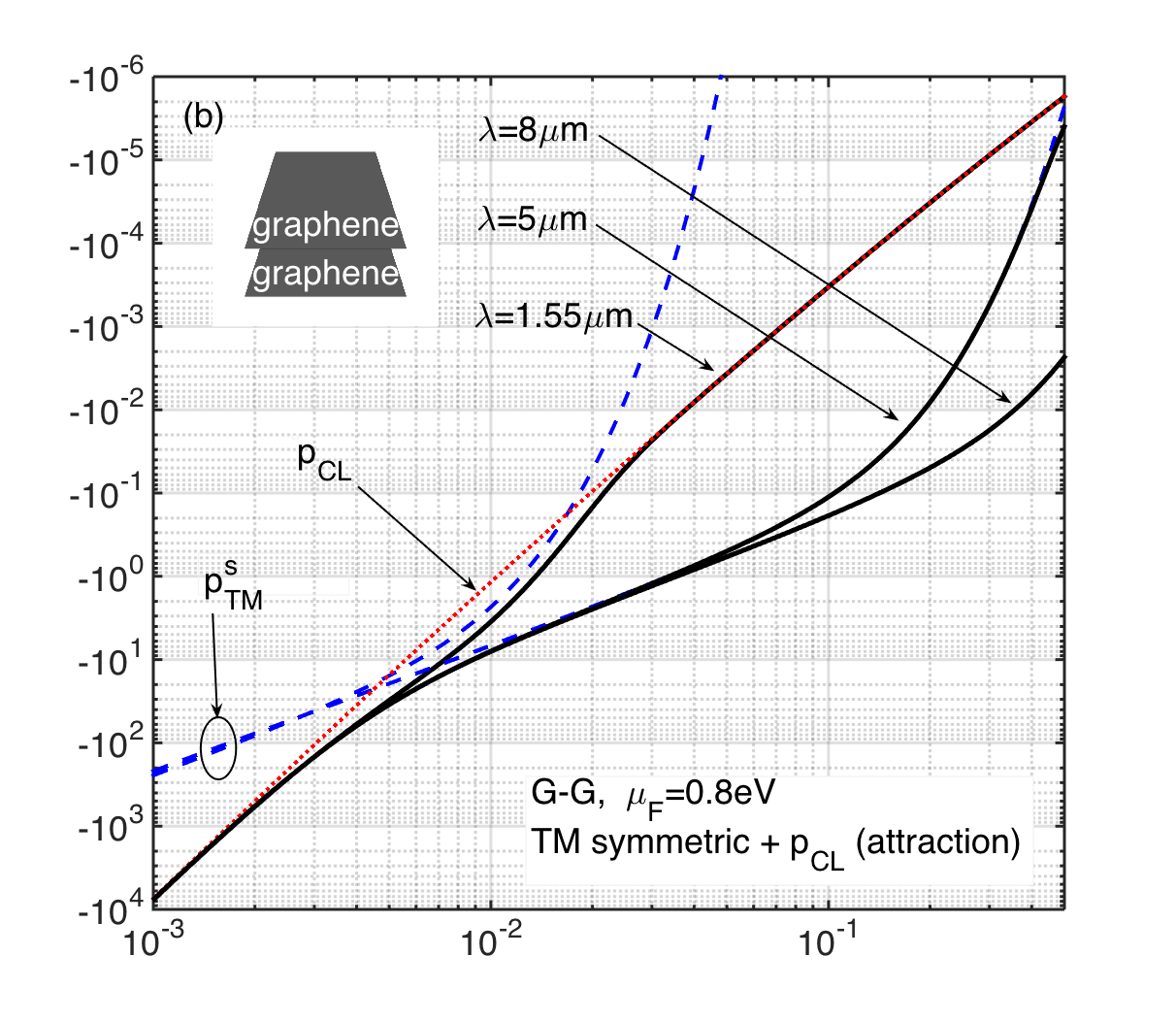}\hspace{-.43cm}
\includegraphics[width=6.25cm,clip=]{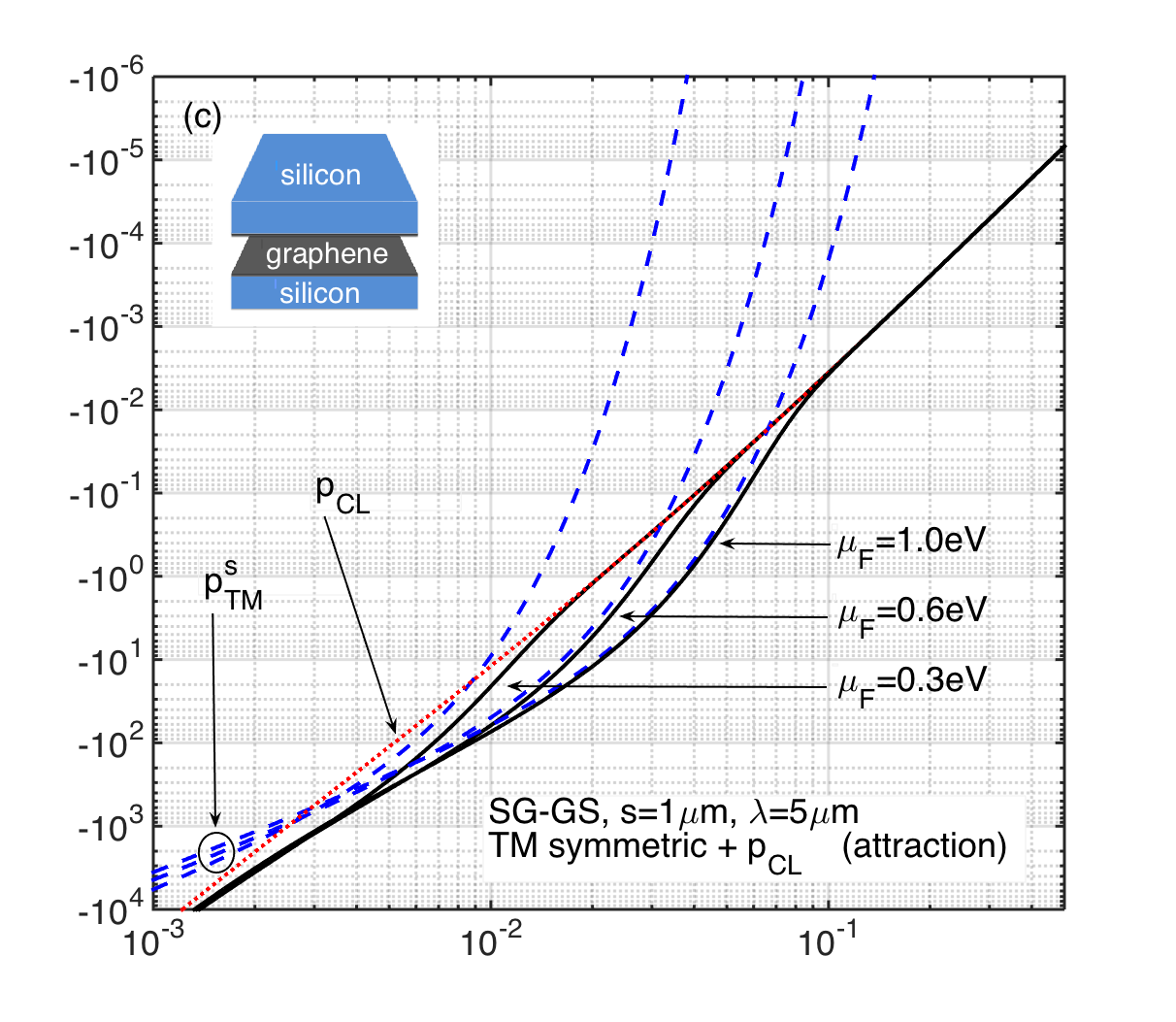}\hspace{-.43cm}
\vspace{-0.2cm}\\
\includegraphics[width=6.25cm,clip=]{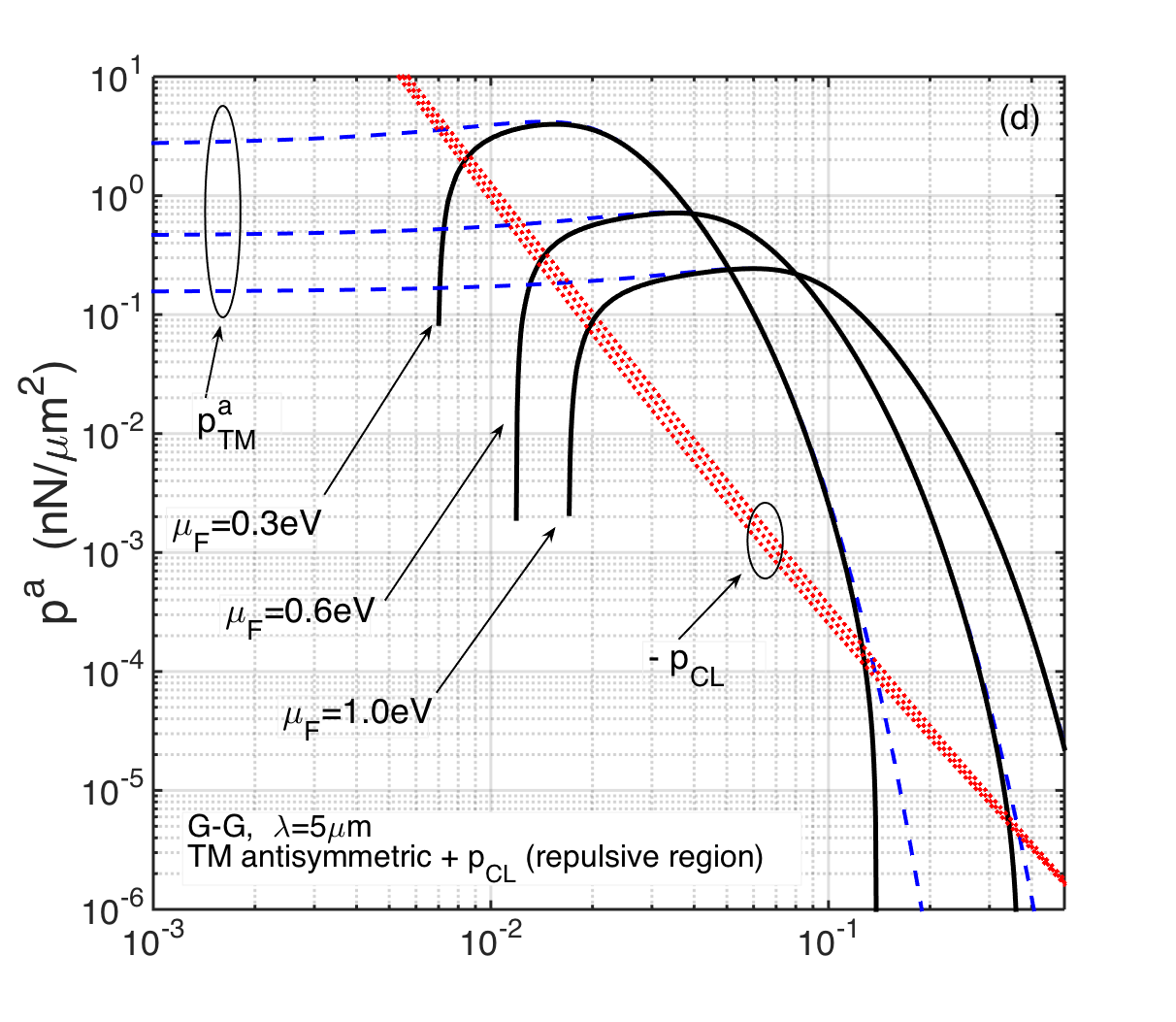}\hspace{-.43cm}
\includegraphics[width=6.25cm,clip=]{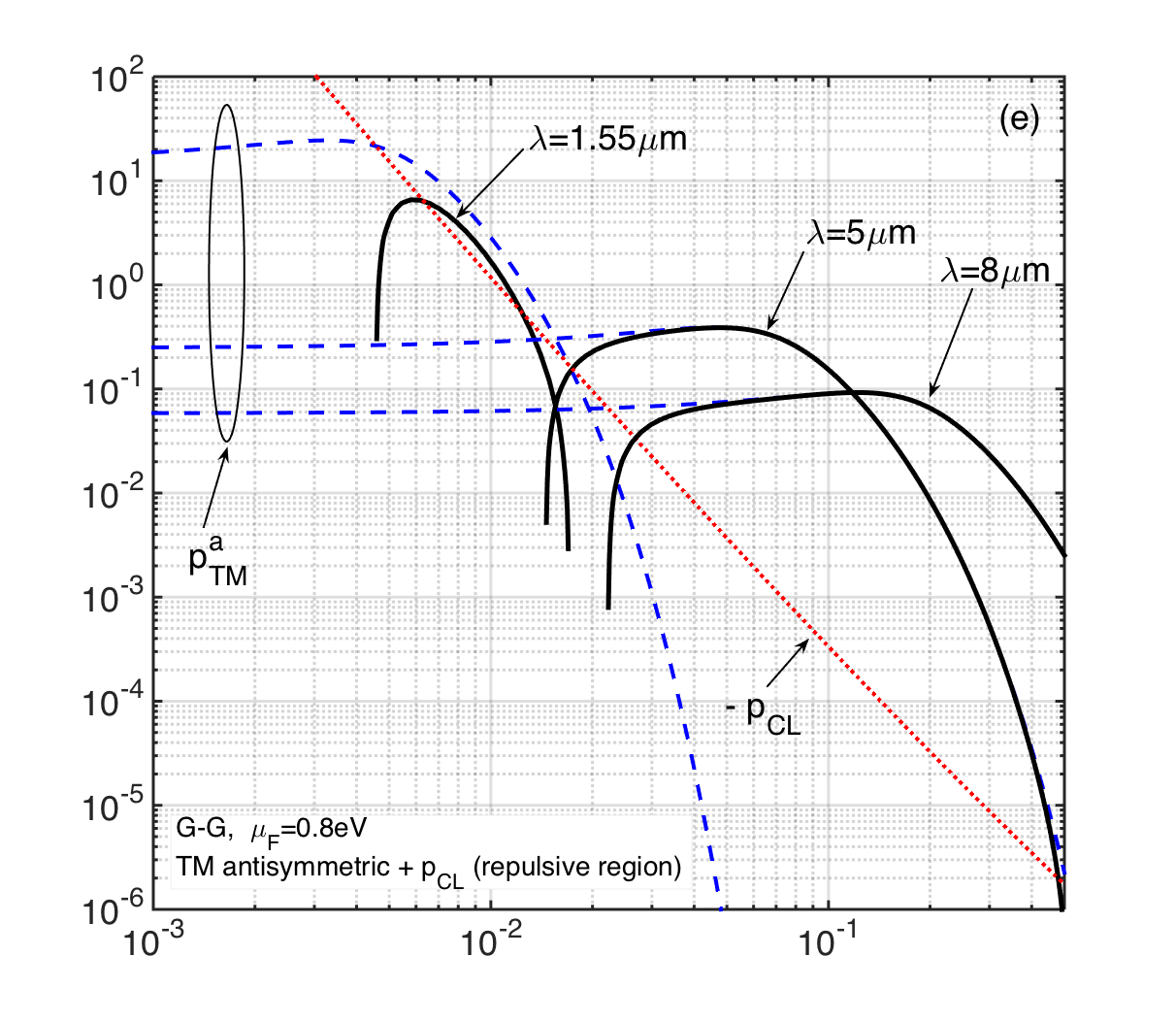}\hspace{-.43cm}
\includegraphics[width=6.25cm,clip=]{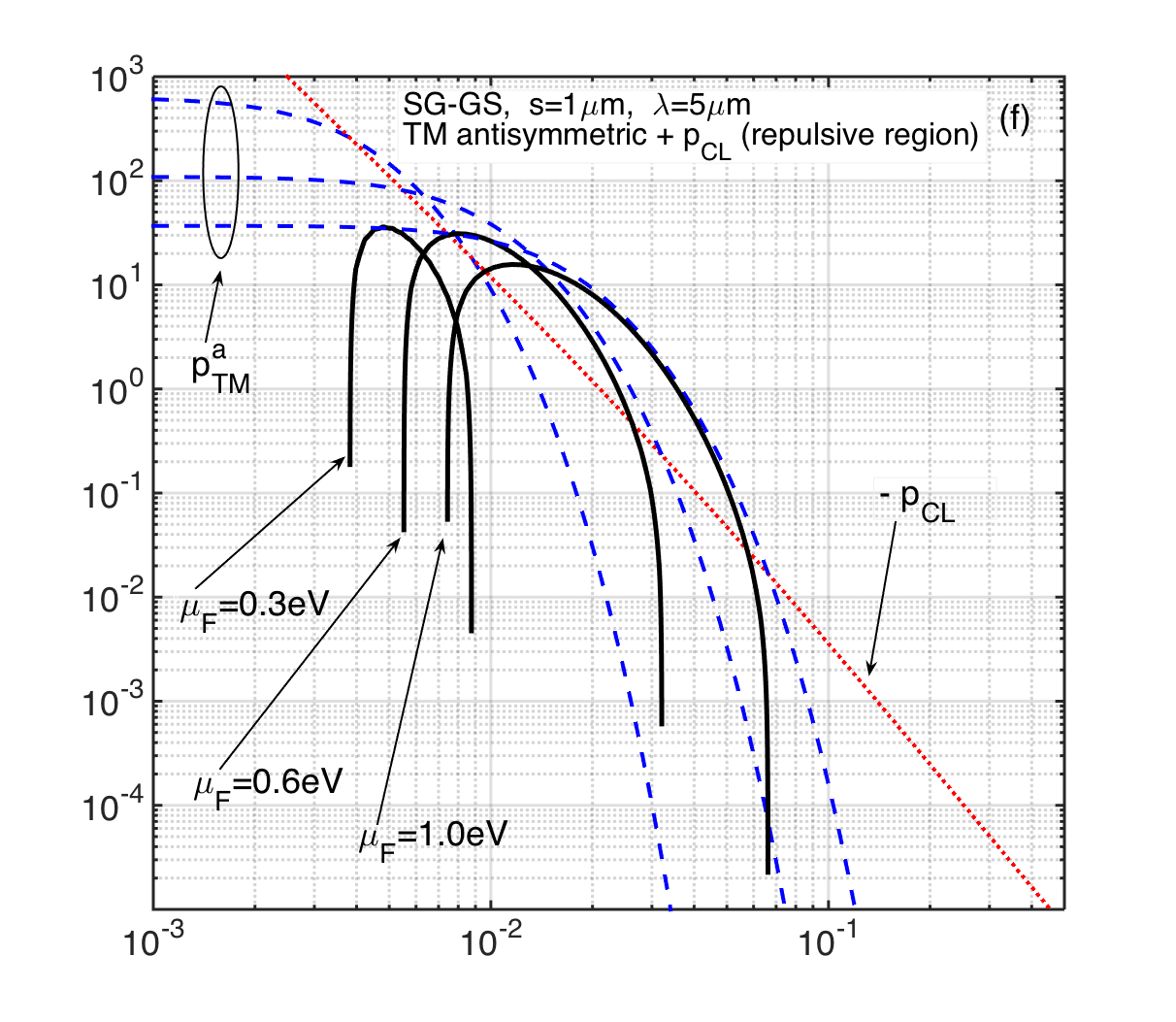}\hspace{-.43cm}
\vspace{-0.2cm}\\
\includegraphics[width=6.25cm,clip=]{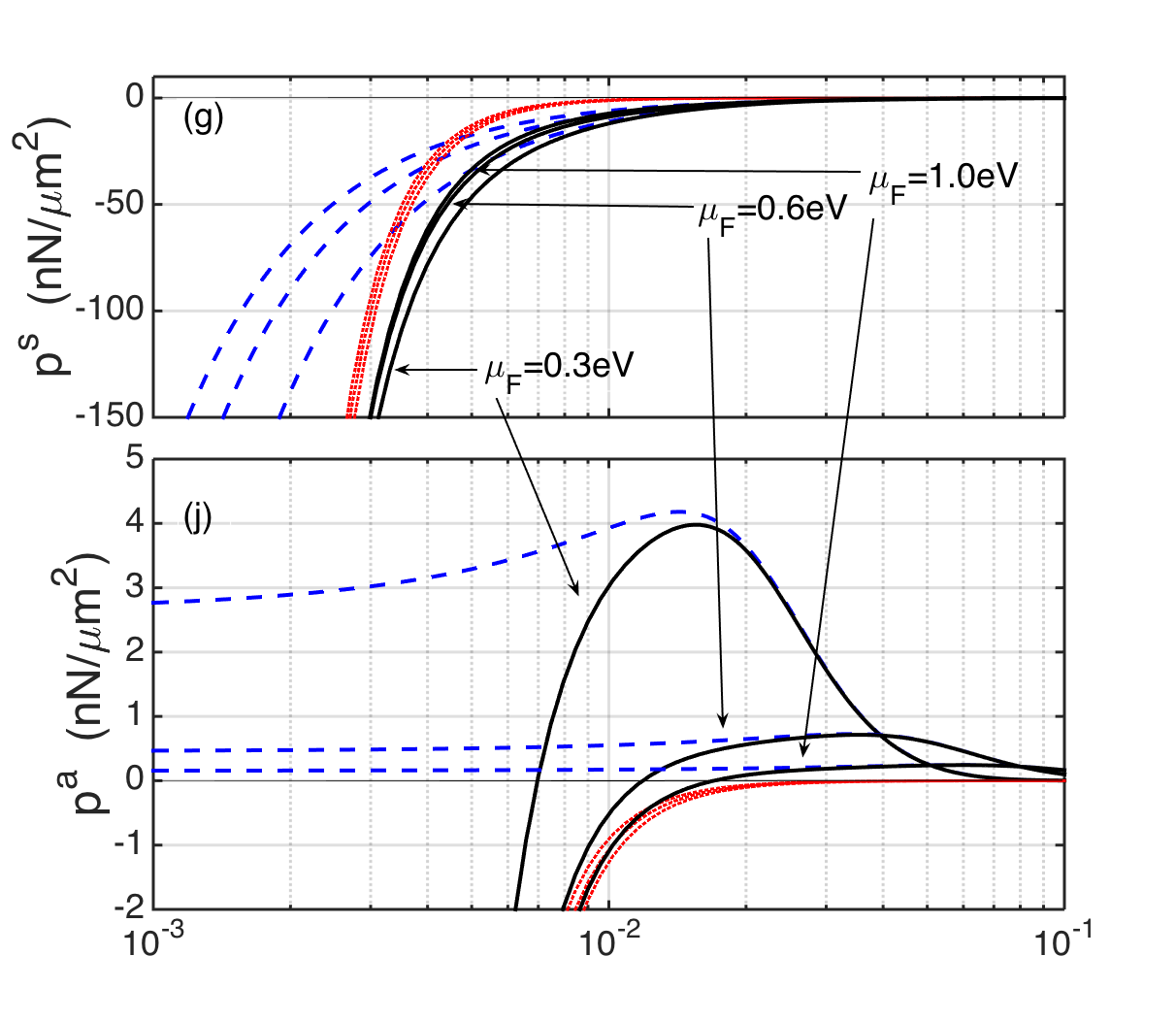}\hspace{-.43cm}
\includegraphics[width=6.25cm,clip=]{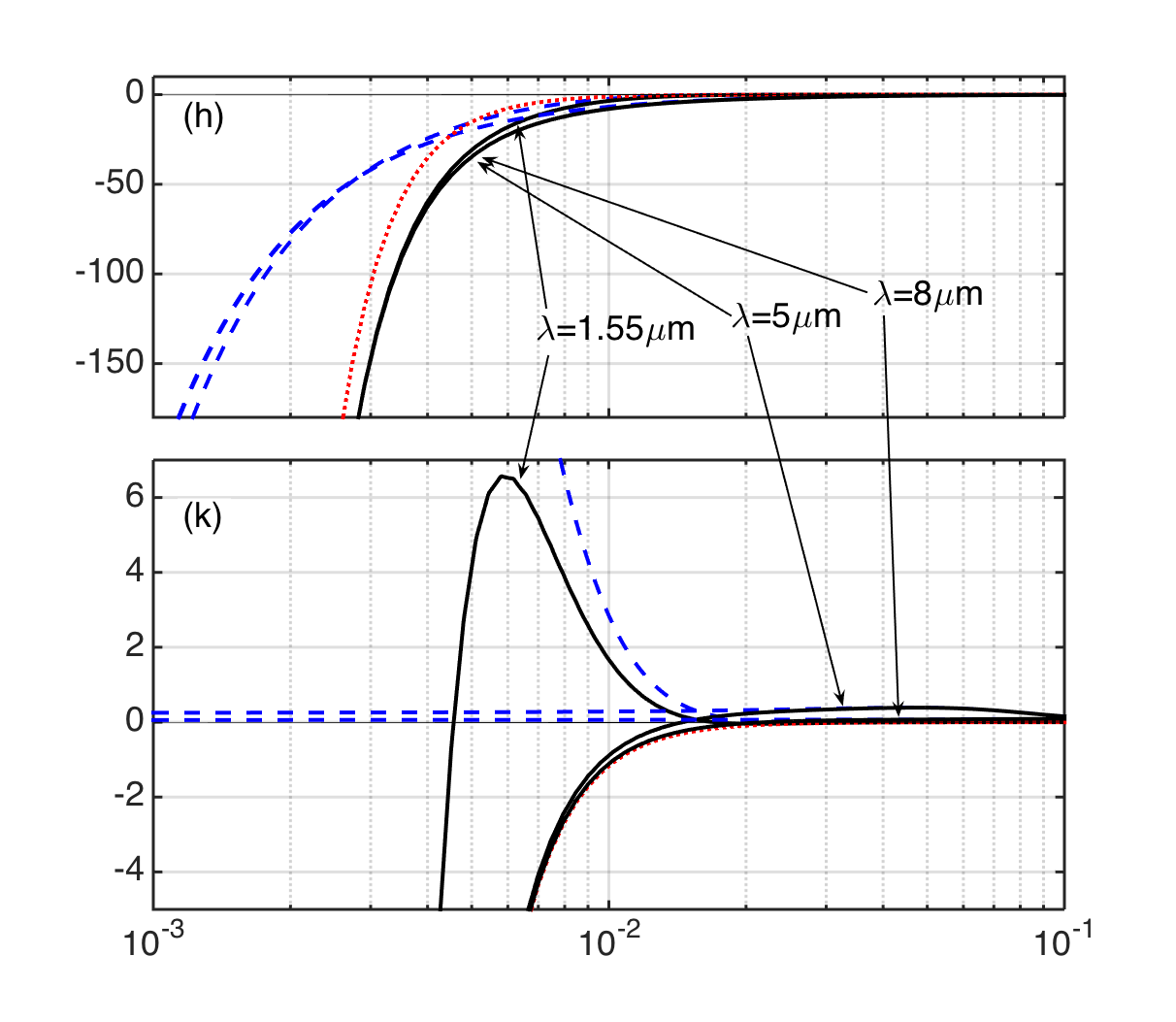}\hspace{-.43cm}
\includegraphics[width=6.25cm,clip=]{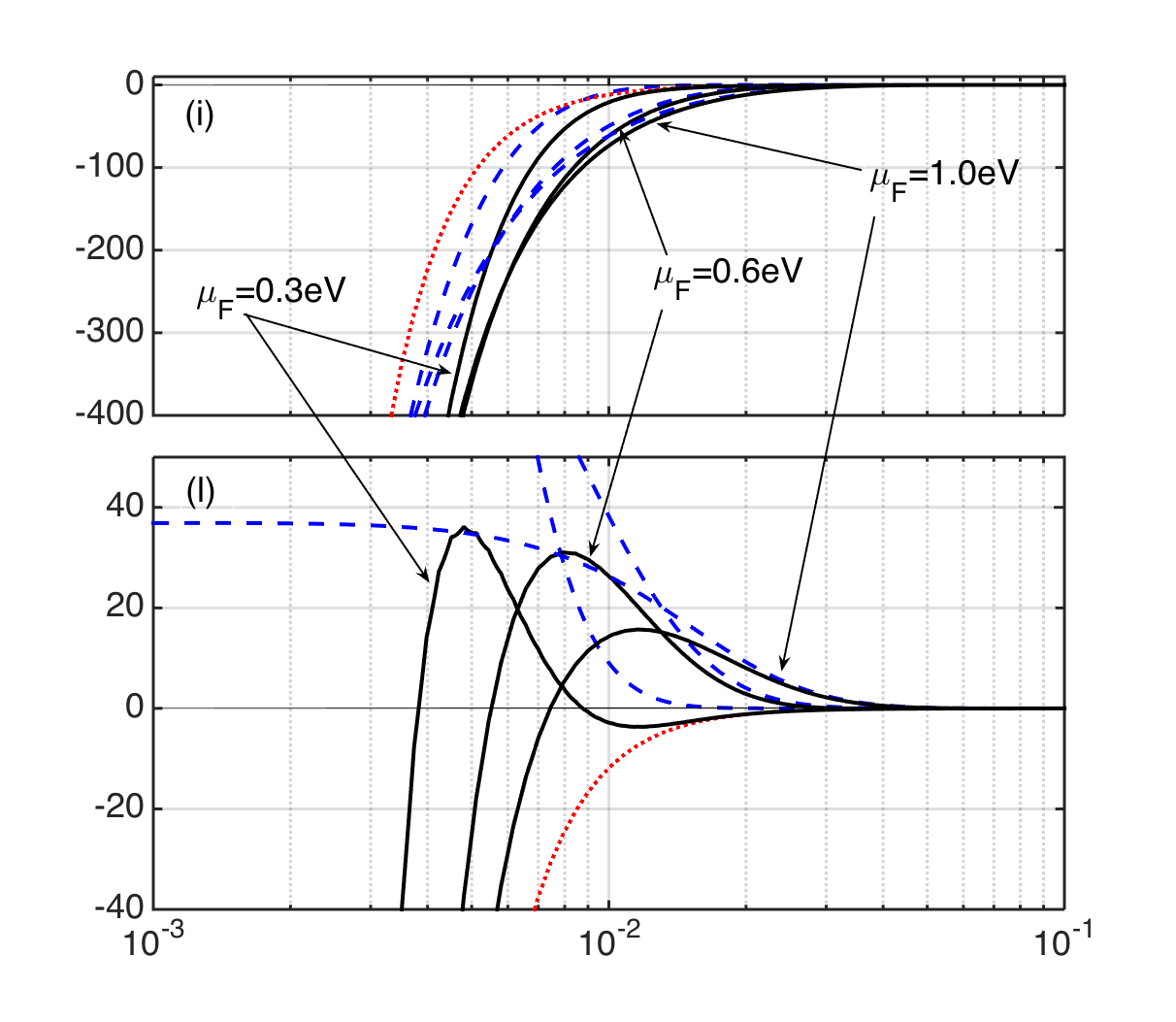}\hspace{-.43cm}
\vspace{-0.2cm}\\
\includegraphics[width=6.25cm,clip=]{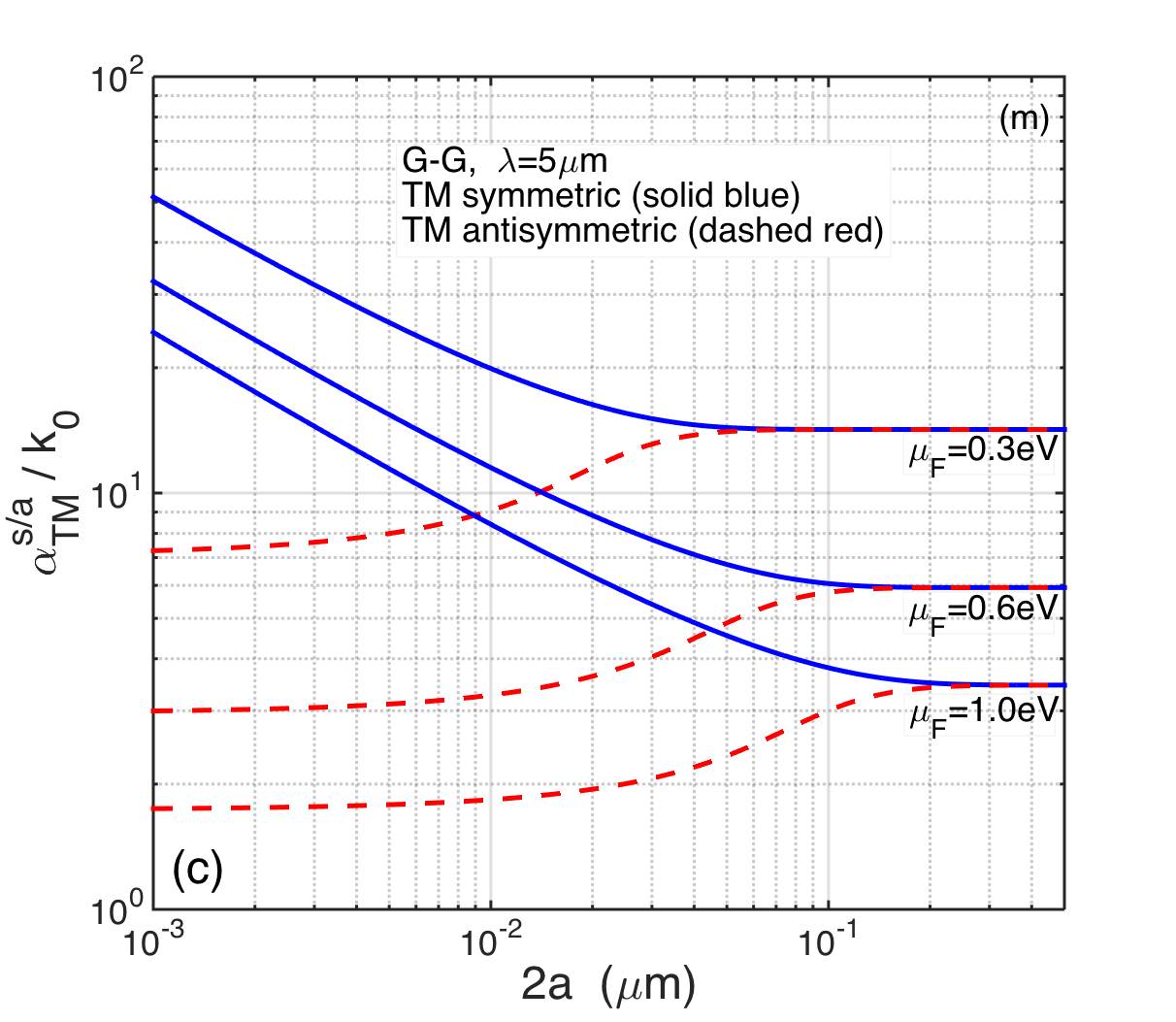}\hspace{-.43cm}
\includegraphics[width=6.25cm,clip=]{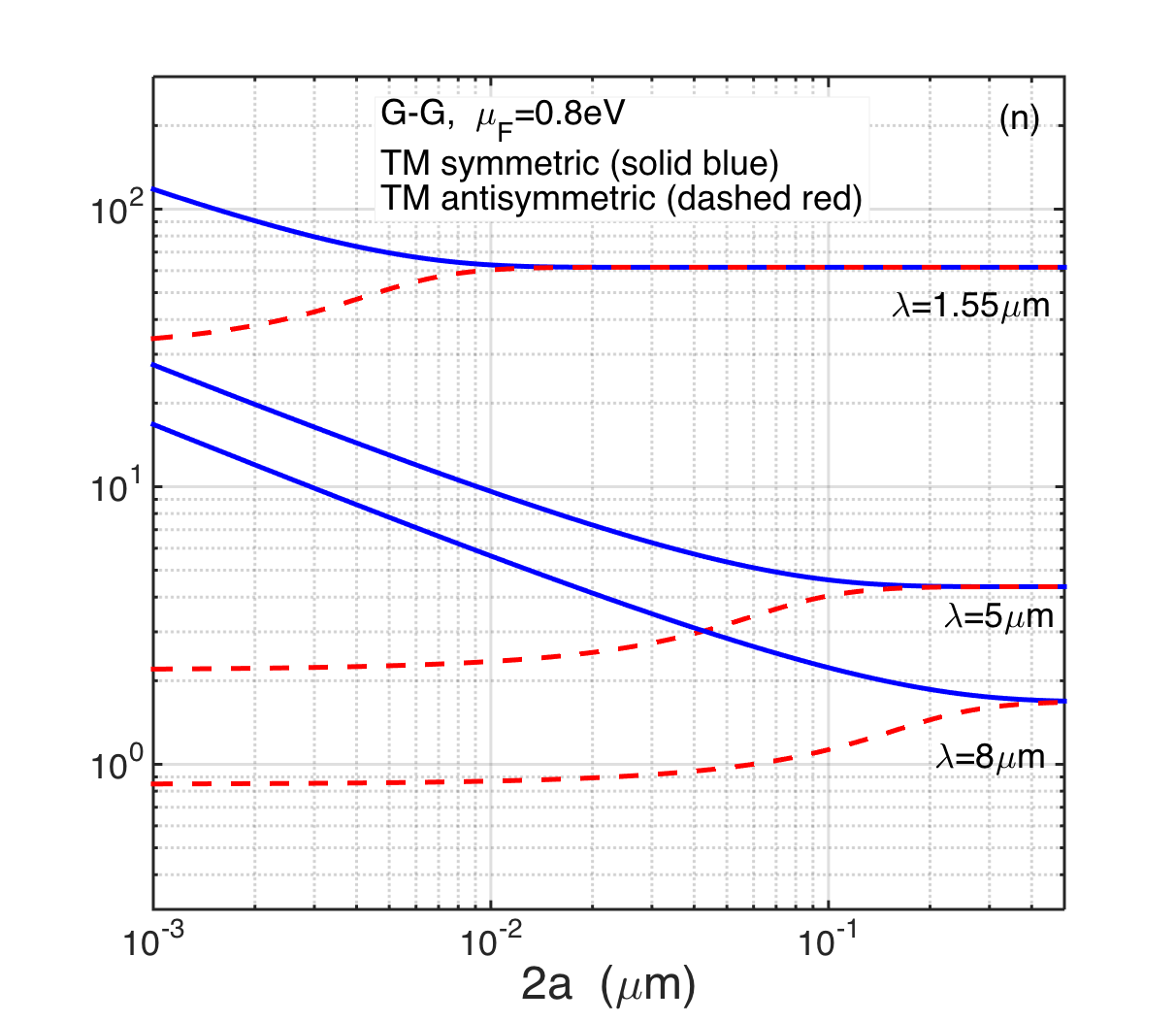}\hspace{-.43cm}
\includegraphics[width=6.25cm,clip=]{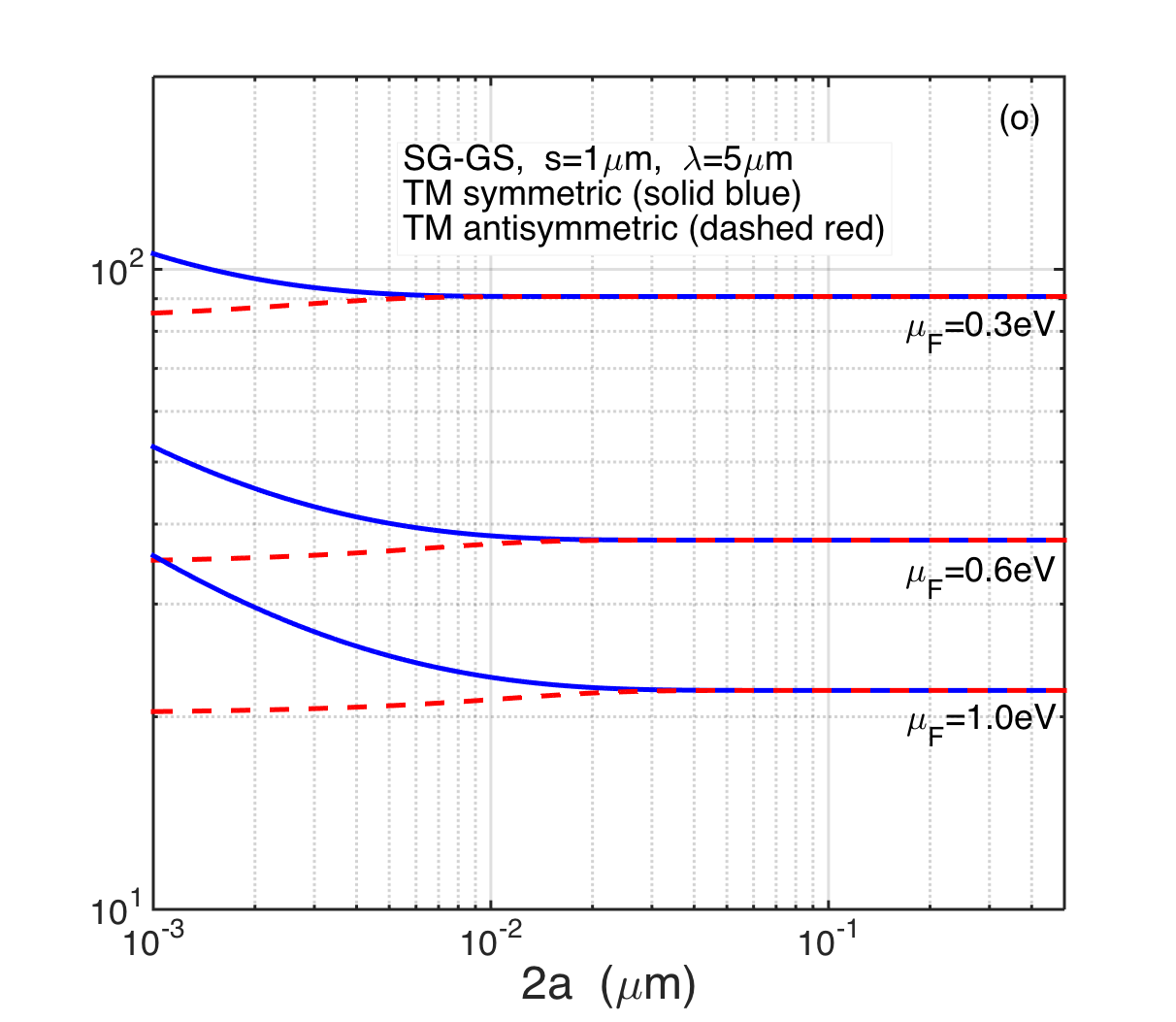}\hspace{-.43cm}
\vspace{-0.5cm}\\
\end{center}
\caption{\label{fig:pressures}\footnotesize (color online). LI and CP pressures, and dispersion relation in region 3, with $T=300$K and $\Gamma=5\;10^{12}$rad/s, and plotted as a function of the separation $2a$. Panels (from a to l): LI (blue dashed) and CL (red dotted, calculated using Eq.~\eqref{eq:Casimir}) pressures, and their sum $p^{s/a}=p^{s/a}_{\textrm{TM}}+p_{\textrm{CL}}$ (black solid) for the TM  symmetric [(a-b-c) log scale and (g-h-i) linear scale]  and antisymmetric [(d-e-f) log scale and (j-k-l) linear scale] modes. Panel (m-n-o): dispersion relation for the TM symmetric (solid blue) and antisymmetric (dashed red) modes, with $\lambda_0=1\mu$m, $\omega_0=2\pi c/\lambda_0$, $k_0=\omega_0/c$. Panels (a-b-d-e-g-h-j-k-n-m): G-G with $\mathcal{P}=1mW/\mu$m [Eq.~\eqref{eq:pressure_TM_gra} for the LI pressure and Eq.~\eqref{eq:ggr3TM} for the dispersion relation]. Panels (c-f-i-l-o): SG-GS with $s=1\mu$m and $\mathcal{P}=20mW/\mu$m [Eq.~\eqref{eq:pressure_TM} for the LI pressure and Eq.~\eqref{eq:TMu3} for the dispersion relation].}
\end{figure*}

Let us first consider modes in the region 3, which are the most interesting.
Figure (\ref{fig:pressures}), in panels from (a) to (l), shows the numerical evaluation of the pressure as a function of the waveguide separation $2a$.  The LI pressure [blue-dashed lines, TM s/a modes, for G-G and SG-GS we used Eq.~\eqref{eq:pressure_TM_gra} and \eqref{eq:pressure_TM} respectively] is plotted for several wavelengths and chemical potentials $\mu_\textrm{F}$ (fulfilling the lossless condition), together with the CL pressure (red-dotted lines), and with their sum (black-solid lines). In the first two columns we considered the G-G configuration, with linear power density $\mathcal{P}=1mW/\mu$m,  while in the third column we considered the SG-GS configuration with linear power density $\mathcal{P}=20mW/\mu$m \cite{nonlinear}.  In the first column of the figure we fixed the mode frequency at $\lambda=5\mu$m ($\omega=0.2 \omega_0$) and varied the graphene chemical potential (see model in section \ref{sec:SlaGra}, with $T=300$K and $\Gamma=5\;10^{12}$rad/s), hence used$\sigma_{\textrm{I}}(\omega, \mu_F=0.3$eV, $0.6$eV, $ 1.0$eV$)=7.47\;10^{-5} ,1.79\;10^{-4},3.08\;10^{-4}$. In the second column of the figure we fixed the graphene chemical potential $\mu_F=0.8$eV and changed the frequency of the mode: $\lambda=8\mu$m, $5\mu$m, $1.55\mu$m corresponding to $\omega/\omega_0=0.125, 0.2, 0645$, respectively, and to $\sigma_{\textrm{I}}(\omega/\omega_0=0.125, 0.2, 0645,\mu_F=0.8$eV$)=3.96\;10^{-4} ,2.44\;10^{-4},5.55\;10^{-5}$. In the third column we made a study with the same graphene parameter used in the first column, but for the SG-GS configuration with Si slab with thickness $s=1\mu$m,  dielectric permittivity $\varepsilon(\omega=0.2\omega_0)=11.7$.

In the logarithmic plots of panels (a) to (f), we recognize the asymptotic behaviors of Eqs.~\eqref{eq:pressyma0}-\eqref{eq:presasyma0} for the LI force. We also see that the CL force dominates at both large and small separations, giving rise to a double change of sign for the antisymmetric pressure [see for instance panels from (d) to (f) and from (j) to (l)]. One of them (the one occurring at larger distances) realizes a position of stable equilibrium [the double change of sign is more pronounced in the case $\mu_{\textrm{F}}=0.3$eV in panel (l)]. Panels from (g) to (l) show, in a linear scale, the same pressures plotted (in a logarithmic scale) in panels from (a) to (f).  

To compare the repulsion with that obtained in other systems we can start by dropping the CL contribution, and evaluate the normalized LI pressure $\beta=p_{\textrm{LI}}dc/\mathcal{P}$, where $d$ is the waveguide separation. The maximum values for LI repulsive pressures are $\beta_{M}\approx6$ for SG-GS, and  $\beta_{M}\approx25$ for G-G configurations. This is one order of magnitudes larger than the state-of-the-art repulsion obtained by nanostructured waveguides  \cite{Oskooi2011} ($\beta_{M}\approx1.9$), and non-structured configurations  \cite{Povinelli2005,Riboli2008} ($\beta_{M}\approx0.1$). The gain of graphene-based waveguides is even larger by considering the attractive CL pressure. Indeed the CL pressure dominates the LI repulsion at small distances, decreasing the maximum attainable repulsion. Remarkably, in the G-G configuration the intensity of the CL interaction is much weaker than in other dielectric or metallic systems  \cite{Rodriguez2011} (typically more than one order of magnitude, several orders for metals), and this allows the LI pressure to dominate down to separations of $\approx10-30$nm, hence attaining a total repulsion which largely overcomes that of other known structures. 

In panels from (m) to (o) we plot the TM dispersion relation for both s/a modes in the G-G and SG-GS configurations [Eq.s~\eqref{eq:ggr3TM} and \eqref{eq:TMu3}, respectively], as a function of the separation $2a$ and of $\mu_\textrm{F}$. We recognize the asymptotic behaviors given by Eq.s~\eqref{eq:allimsym} and \eqref{eq:allimasym}. 

%

\begin{figure}[htb]
\includegraphics[width=0.48\textwidth]{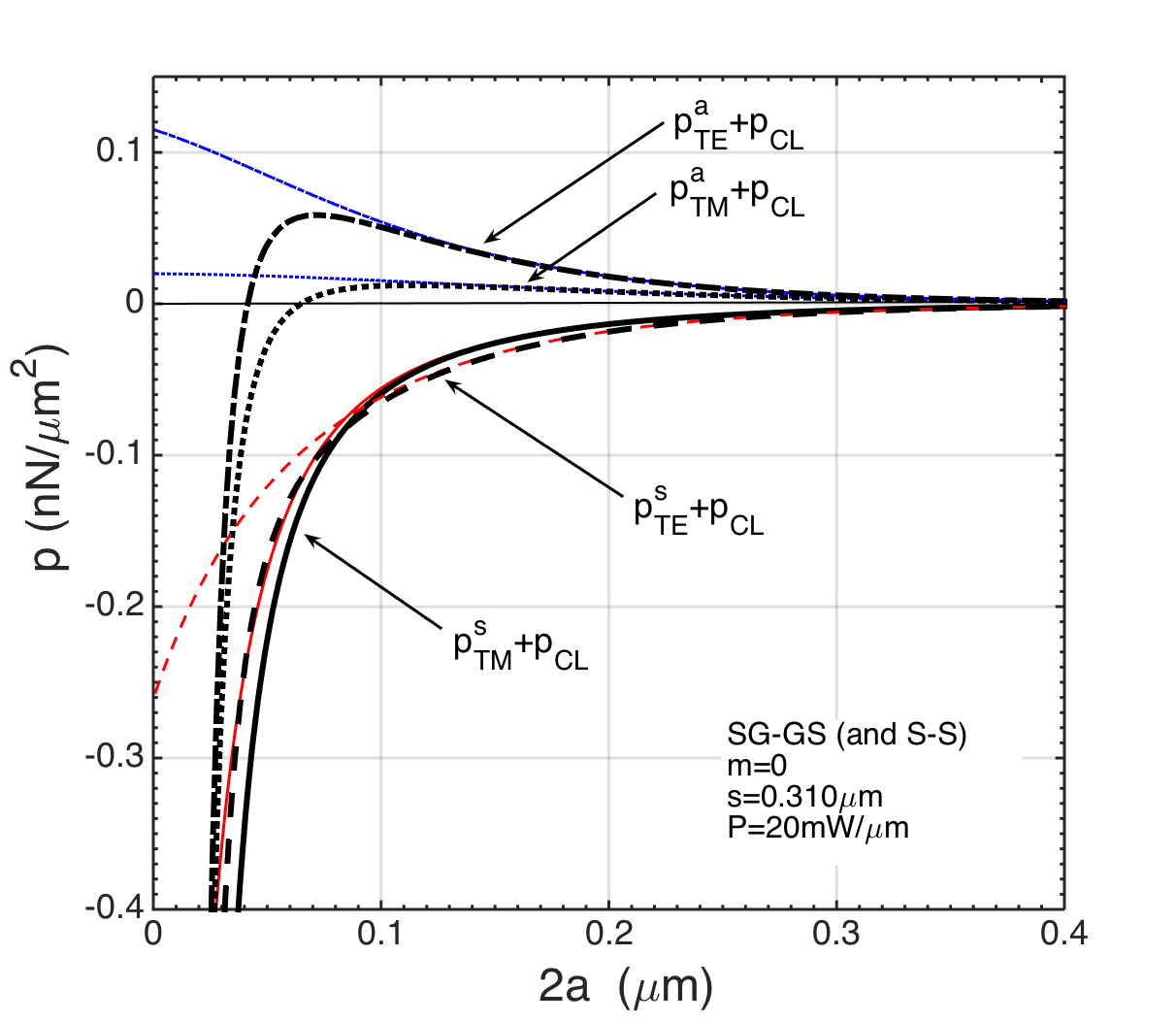}
\caption{\label{fig:fig_press_SS_s0310_P20}\footnotesize (color online). LI and CP pressures in region 2 for the SG-GS and S-S configurations. Silicon slabs are of thickness $s=0.310\mu$m, $\Gamma=5\;10^{12}$rad/sec, $T=300K$. The plotted lines remain unchanged for the different values of $\mu_F=0.3$eV, $0.6$eV, $1.0$eV. Total radiation pressure $p^{s/a}_{TE/TM}+p_{CL}$ for the symmetric TE and TM modes (black dashed and solid lines respectively)  and for the antisymmetric TE and TM modes (black dash-dotted and dotted lines, respectively). $p_{CL}$ has been calculated using Eq.~\eqref{eq:Casimir}.  The LI pressures $p^{s}_{TE}$ (red dashed), $p^{s}_{TM}$ (red solid), $p^{a}_{TE}$ (blue dash-dotted), $p^{a}_{TM}$ (blue dotted), are calculated using  Eq.s~\eqref{eq:pressure_TE} and \eqref{eq:pressure_TM}, with $\lambda=1.55\mu$m, $\varepsilon_{\textrm{R}}=12.11$, $m=0$, and $\mathcal{P}=20mW/\mu$m. These curves remains unchanged even by eliminating the graphene sheets [i.e. S-S configuration, $\sigma=0$ in Eq.s~\eqref{eq:pressure_TE} and \eqref{eq:pressure_TM}].}
\end{figure}

Finally let us consider the pressure corresponding to modes in the region 2. In figure (\ref{fig:fig_press_SS_s0310_P20}) we plotted both the LI [ Eq.s~\eqref{eq:pressure_TE} and \eqref{eq:pressure_TM}] and CL pressure [Eq.~\eqref{eq:Casimir}] for the SG-GS configuration with $s=0.310\mu$m for the modes TE/TM s/a with $m=0$ (see more details in the figure caption). We remark that the curves do not change by varying $\mu_F$ and also by completely eliminating the graphene sheets.  Hence the light induced pressures $p^{s/a}_{TE/TM}$ correspond to that of Fig. 3 of  \cite{Riboli2008} [here they are 10 times weaker due to a typo in  \cite{Riboli2008}]. We see that the role played by the CL force is important, and cannot be neglected. 

\section{Conclusions}\label{sec:conclusions}
We studied the light-induced forces occurring in graphene-based (suspended or supported) optomechanical waveguides. We derived the dispersion relations, the relevant device length-scales and the explicit analytical closed form expression of the LI forces. While for dielectric or metallic waveguides the LI pressure is always bounded, in presence of graphene the TM symmetric mode dispersion relation diverges as $1/a^{1/2}$ at small separations $2a\to 0$, implying an attractive force diverging as $-1/a^{3/2}$.  We also calculated the additional fluctuation-induced Casimir-Lifshitz force, which is always attractive and dominates at short and large distances (it can dominate over the repulsive TM asymmetric mode both at small and large separations, giving rise to a position of stable equilibrium). Thanks to a combined effect of a strong field confinement with a weak CL attraction, the total force is considerably stronger than for the most optimized complex nanophotonic structures. It is widely tunable by varying the chemical potential via chemical or via a simple electrostatic doping, allowing for a fast modulation. These features open a new path for micro- nano-scale sensors and optomechanical devices based on graphene and other 2D materials  \cite{2D}.

\appendix

\section{Light-induced electromagnetic force }\label{sec:PhySysEleFor}
In order to calculate the time-averaged optical force $F$ induced by the excited light mode of the structure and acting on part of the systems (let us say the graphene sheet and its supporting slab in the positive z half-space) one should evaluate the surface integral \cite{Jackson,LL}:
\begin{equation}\label{eq:force}
{\bf F}=\int_{\Sigma} {\bf T}({\bf r}) \cdot {\bf n}\;d\sigma
\end{equation} 
where $\Sigma$ is a closed oriented surface enclosing the object (in vacuum) on which the force is to be evaluated, ${\bf n}$ is the unit vector normal to the surface, and ${\bf T}=\langle\mathbb{T}({\bf r},t)\rangle_t$ is the time averaged Maxwell stress tensor in vacuum whose components are 
\begin{multline}\label{eq:MaxST}
\mathbb{T}_{ij}({\bf r},t)=\varepsilon_0 \left[ e_ie_j+(\mu_0c)^2\;h_i h_j \right.\\
\left.-\frac{1}{2}\left(e^2+(\mu_0c)^2\;h^2\right)\delta_{ij}\right],
\end{multline}
where $c$ is light velocity, $\mu_0$ and $\varepsilon_0$ being respectively the vacuum permeability and permittivity. For a monochromatic electromagnetic field ${\bf e}(\rr,t)=\Rea\left[{\bf E}(\rr)\;e^{-i\omega t}\right]$ and ${\bf h}(\rr,t)=\Rea\left[{\bf H}(\rr)\;e^{-i\omega t}\right]$, ${\bf E}(\rr)$ and ${\bf H}(\rr)$ are $\omega$ dependent, and $\langle e_i(\rr,t)e_j(\rr,t)\rangle_t=\textrm{Re}[E_i(\rr)E_j^*(\rr)]/2$.

For symmetry reasons and in the absence of losses  \cite{Antezza2011,Riboli2008} the force acts only in the $z$ direction, the only contributing component of the Maxwell stress tensor is $T_{zz}$, the maxwell stress tensor is uniform in the $xy$ plane, hence the pressure acting on the upper graphene-slab bilayer is:
\begin{equation}\label{eq:force_z_noloss}
p_{\textrm{LI}}=\frac{F_z}{\Sigma_3}=T_{zz}({\bf r}\in \Sigma_5) -T_{zz}({\bf r}\in \Sigma_3) ,
\end{equation} 
where 
\begin{multline}\label{eq:T_zz}
T_{zz}=-\frac{\varepsilon_0}{4}\left[|E_x|^2+|E_y|^2-|E_z|^2+\right.\\
\left.\mu_0^2c^2\left(|H_x|^2+|H_y|^2-|H_z|^2\right)\right],
\end{multline} 
and $\Sigma_5$ ($\Sigma_3$) is a parallel plane over (below)  of the graphene-slab bilayer. Once the fields are known (see section \ref{sec:DispRel}) one can show  \cite{Riboli2008} that $T_{zz}({\bf r}\in \Sigma_5)=0$, hence obtaining Eq.~\eqref{eq:LIpress}.


\end{document}